\documentclass[a4paper,11pt]{article}
\usepackage{jheppub}
\usepackage[T1]{fontenc}
\usepackage{cancel}
\usepackage{amsmath}
\usepackage{slashed}
\usepackage{graphicx}
\usepackage{multicol}
\usepackage{footnote}
\usepackage{subfigure}
\usepackage{float}
\usepackage[utf8]{inputenc}
\usepackage{color}
\DeclareGraphicsExtensions{.eps}
 \usepackage[normalem]{ulem}
\long\def\/*#1*/{}
\usepackage{color}
 \usepackage[normalem]{ulem}
 
\definecolor{darkgreen}{cmyk}{1,0,1,0.4}

\def\barr{\begin{array}}
\def\earr{\end{array}}
\def\dis{\displaystyle}
\def\mev{\, {\rm MeV}}
\def\gev{\, {\rm GeV}}

\def\lapp{\mathrel{\rlap{\raise.5ex\hbox{$<$}}
                    {\lower.5ex\hbox{$\sim$}}}}
\def\gapp{\mathrel{\rlap{\raise.5ex\hbox{$>$}}
                    {\lower.5ex\hbox{$\sim$}}}}

\title{\boldmath Model Independent analysis of MeV scale dark matter: II.
  Implications from $e^-e^+$ colliders and Direct Detection}

\abstract{Dark matter particles with masses in the sub-GeV range
  have escaped severe constraints from direct detection experiments
  such as LUX, PANDAX-II and XENON100 as the corresponding recoil
  energies are, largely, lower than the detector thresholds. In a
  companion paper, we demonstrated, in a model independent approach,
  that a significantly large fraction of the parameter space escapes
  the cosmological and astrophysical constraints. We show here, though,
  that the remaining parameter space lends
  itself to the possibility of discovery at both direct detection
  experiments (such as CRESST-II) as well as in a low-energy collider
  such as Belle-II.  \\[1ex]
  }

\author[a]{Debajyoti Choudhury,}
\author[a,1]{Divya Sachdeva\note{Corresponding author (divyasachdeva951@gmail.com)}}

\affiliation[a]{Department of Physics and Astrophysics, University of Delhi, Delhi 110 007, India}

\emailAdd{debajyoti.choudhury@gmail.com}
\emailAdd{divyasachdeva951@gmail.com}

\begin{document}
\maketitle
\section{Introduction}
   \label{sec:intro}
The evidence for Dark Matter, at least as far as the manifestations of
its gravitational interactions are concerned, has been continuously
building up. Whether it be rotation curves in spiral
galaxies~\cite{Rubin:1970zza}, the observation of gravitational
microlensing ~\cite{Moustakas:2002iz,2012A&A...545A..71V},
observations of cluster collisions (Bullet
Cluster)~\cite{Clowe:2006eq}, or the temperature anisotropy in the
spectrum of Cosmic Microwave Background
Radiation~\cite{Hinshaw:2012aka,Abadi:2002tt,Ade:2015xua}, there exist
a large class of observations, spanning very different length scales,
for which the Dark Matter (DM) hypothesis provides the most compelling
explanation. And while efforts to circumvent particulate DM have been
made, primarily through modifications of Einsteinian
gravity at cosmological scales~\cite{Milgrom:2001ny,Famaey:2011kh},
neither can a single such modification explain all data, nor are
theories incorporating such modifications necessarily unrelated from
a model involving particles as DM~\cite{Calmet:2017voc}.

On the other hand, no direct ({\em i.e.}, laboratory) evidence for such
DM particles has been forthcoming despite a large variety
of experiments having been operative. All such efforts hinge upon the
assumption that the DM would have some interaction with the Standard
Model (SM) particles\footnote{Such an assumption also facilitates the
  attainment of the correct relic density.}. Such search strategies can
be broadly categorized into three classes, namely \textit{a)} satellite
based {\em indirect detection} experiments like
Fermi-LAT~\cite{Ackermann:2015zua}, PAMELA~\cite{Adriani:2008zr} and 
AMS~\cite{Lin:2015taa}, \textit{b)} specialized terrestrial {\em
  direct detection experiments} and \textit{c)} generic {\em collider}
experiments.  Despite occasional claims of anomalies in the data,
putative positive sightings have never been validated by a different
experiment, thereby leading to progressively stronger constraints on
the parameter space of any theory of DM.

Most of the aforementioned search strategies have concentrated on a
relatively heavy ({\em i.e.}, heavier than a few GeVs) DM particle.
Indeed, indirect search experiments depend upon the annihilation of a
pair of DM particles into SM particles, leading to aberrant cosmic
rays (such as those generated by antiparticles like
positrons or anti-protons), very high energy neutrinos, monochromatic
photons or even an anomalous component of the continuous $\gamma$-ray
spectrum. Corresponding particles from the annihilation of light DM
particles would have energies typically well below the threshold of
current satellite-based detectors. Similarly, in direct detection
experiments, the scattering of a light DM-particle off the target
nuclei would, typically, impart too little a recoil to the latter to
be distinguishable above the background (due to both thermal
fluctuations as well as the scattering of the ambient
neutrinos). Entirely analogous arguments would hold for, say, the
Large Hadron Collider, where the associated production of such DM
particles would lead to a relatively small recoil of the visible
particle system, with the consequent missing transverse energy
spectrum being hardly recognizable from that due to neutrinos
(appearing in corresponding events in the SM background). In short,
sub-GeV DM affords a much larger room as far as the constraints from
canonical experiments are concerned.

This, as well as several other theoretical compulsions have engendered
much recent interest in sub-GeV DM
particles~\cite{Hui:2016ltb,Alexander:2016aln,Heeck:2018lkj,
  Matsumoto:2018acr,Feng:2008ya}. In particular, towards the
explanation of perceived anomalies in the 511 keV $\gamma$-rays
observed by the INTEGRAL satellite, the cosmic $\gamma$-ray background
at 1-20 MeV and the details of large scale structure, quite a few such
models~\cite{Boehm:2003bt,Ahn:2005ck,Borodatchenkova:2005ct,Hooper:2007tu}
have been invoked over the years.  The wide plethora of physics
scenarios that can, generically, lead to such ultralight particles
makes the subject a very fascinating one. And since standard methods
do not work, the exception being those emanating from anomalous decays
of certain mesons~\cite{Li:2018hgu}, new methods need to be devised
for exploring these. Indeed, quite a few diverse ideas have already
been proposed, such as the absorption signal in the 21-cm
spectrum~\cite{Barkana:2018qrx,Barkana:2018cct}, the scattering of DM
off atomic clocks~\cite{Wolf:2018xlz}, the use of optical
cavities~\cite{Geraci:2018fax}, the use of leptonic
beam-dumps~\cite{Marsicano:2018glj}, the use of a cryogenic
point-contact germanium detector~\cite{Jiang:2018lij} or more
canonical setups  such as the LDMX~\cite{Geraci:2018fax}.

In this paper, we examine, instead, the viability of searching for
such light DM in an existing collider facility,
namely Belle-II\footnote{While analogous studies have been performed
  earlier, both at the phenomenological
  level~\cite{Borodatchenkova:2005ct} as well as a full experimental
  search~\cite{Seong:2018gut}, these have been in the context of
  specific models unlike in our approach.}.  In a companion
paper~\cite{Choudhury:2019tss}, hereafter designated Paper I, we have
examined the cosmological constraints on such a DM paradigm. Here, we
consider, primarily, the sensitivity reach for a host of different
final states at Belle-II. The relatively low
energy, the clean environment, and the high luminosity all work in our
favour. While the insistence on low-energy might seem
counterintuitive, we explicitly show the advantage thereof by
comparing with the reach that would have been possible at LEP. Also
considered are the prospects of Direct Detection experiments.

\section{Higher Dimension operators}
\label{sec:operators}
Rather than consider an intricate and 
ultraviolet-complete model, we take recourse to a 
model-independent approach, 
with the only assumption being that the light DM candidate $\varphi$ is a spin-0 
particle. While the effective field theory approach pertaining to our case 
has been detailed in Paper I~\cite{Choudhury:2019tss}, we recount this here 
for the sake of completion. With the mediator 
connecting the dark sector to the SM particles considered to be heavy enough to 
be integrated out\footnote{Similarly, any other new species is 
also assumed to be too heavy to be relevant 
in the contexts of both terrestrial experiments/observations
as well as the cosmological evolution of the
relic density.}, the only new relevant field is the
scalar. Since we are interested in a DM with a mass of at most
a few GeVs, the only relevant SM states are the photon and the gluon,
the leptons (including neutrinos) and the quarks of the first two
generations. Furthermore, flavour changing operators are omitted 
so as to be trivially consistent with low-energy constraints.

Assuming $SU(3) \otimes U(1)_{\rm em}$ symmetry\footnote{Had we
  imposed the full gauge symmetry of the SM instead, the first two
  operators, viz. $\mathcal{O}^f_{s,p}$ would suffer a further
  suppression by a factor of $v / \Lambda$ where $v$ is the
  electroweak symmetry breaking scale. We return to this point
    later.}, the lowest-dimensional operators are
\begin{equation}
\begin{array}{lcl}
\mathcal{O}^f_{s}&=& \dis \frac{\mathcal{C}^f_{s}}{\Lambda} \varphi^\dagger\varphi
   \; \bar{f} f \\[2ex]
\mathcal{O}^f_{p}& = &\dis \frac{\mathcal{C}^f_{p}}{\Lambda} \varphi^\dagger\varphi \,
         \bar{f}\gamma^5 f
\\[2ex]
\mathcal{O}^f_{v}& = &\dis \frac{\mathcal{C}^f_{v}}{\Lambda^2}i~( \varphi^\dagger\partial_{\mu}\varphi-\partial_{\mu}\varphi^\dagger \varphi) \; \bar{f}\gamma^{\mu} f 
\\[2ex]
\mathcal{O}^f_{a}& = &\dis \frac{\mathcal{C}^f_{a}}{\Lambda^2}i~( \varphi^\dagger\partial_{\mu}\varphi-\partial_{\mu}\varphi^\dagger \varphi) \; \bar{f}\gamma^{\mu}\gamma^5 f
\\[2ex]
\mathcal{O}_{\gamma}& = &\dis \frac{\mathcal{C}_{\gamma}}{\Lambda^2}( \varphi^\dagger \varphi) F_{\mu\nu} F^{\mu\nu}	
\\[2ex]
\mathcal{O}_{\tilde{\gamma}}& = &\dis \frac{\mathcal{C}_{\tilde{\gamma}}}{\Lambda^2}( \varphi^\dagger \varphi) F_{\mu\nu} \tilde{F}^{\mu\nu} \ ,	
\end{array}
\label{the_operators}
\end{equation}
where $f$ is an arbitrary SM fermion and $\Lambda$ is the scale of new
physics. Note that the first two operators are dimension-5 ones
  while the rest are dimension-6. This difference would manifest
  itself in the experimental sensitivities. With the dimensionless Wilson coefficients ${\cal C}$'s, corresponding to the various
operators, being normalized to either zero or unity
(denoting the absence or presence of the said operator), the results
would be functions of the mass of DM and the scale $\Lambda$ alone. 
The translation to the parameter space of a 
UV-complete theory would, then, be a straightforward one.

\section{Monophoton signal at Belle-II}
   \label{sec:monophot_belle}
In this section, we explore the sensitivity of the low-energy $e^-
e^+$ collider Belle-II to such a DM candidate.  The relatively low
center-of-mass energy ($\sqrt{s}$\,=\,10.58\,GeV), alongwith the high
luminosities available (1--50$\,{\rm ab}^{-1}$) in this experiment,
renders Belle-II a very attractive theatre for the search of such
light DM particles. It is instructive to examine this contention
carefully.  Purely on dimensional grounds, a typical cross section of
interest driven by either of the first two operators in
eq.(\ref{the_operators}) would scale, with the center-of-mass
  energy, as $\Lambda^{-2} \, \ln(s/m_e^2)$. Similarly, those driven by the other
operators in eq.(\ref{the_operators}) would scale as $s \,
\Lambda^{-4} \, \ln(s/m_e^2)$. On the other hand, the various
components of the SM background would, naively, be expected to fall as
$s^{-1} \ln (s/m_e^2)$ or even faster.  Thus, a larger center-of-mass
energy would, seemingly, serve to increase the signal to background
ratio. This is more than offset, though, by the nature of the signal
and background. With the DM being stable and largely noninteracting,
the signal final state would comprise of a visible particle
accompanied by missing energy-momentum.  The latter, within the SM,
accrues primarily from neutrino-production (apart from the
experimental effect\footnote{As we shall see later, this instrumental
  background tends to overwhelm that from neutrino production.} of
having missed ostensibly visible particles). The corresponding rates
fall dramatically as $\sqrt{s}$ falls well below $M_Z$, and, in the
regime of interest, would scale as $G_F^2 \, s \, \ln(s/m_e^2)$. In
other words, the energy-dependence of such background is the same as
that for the dimension-6 operators, and, potentially, worse than that
for the dimension-5 ones. This is what renders an experiment such as
Belle a very interesting arena for the search of light DM candidates.

While, at a given collider, the DM particle can be produced in many
different processes, only a few of them are, potentially, of interest.
With the DM particle being produced only in pairs, there must be at
least one visible particle in the final state for the event to be
triggered. The simplest of such processes is where a single photon is
emitted alongwith the pair of DM particles, {\it viz.}
\begin{equation}
\label{signal_process}
e^+ e^- \to \varphi^* + \varphi + \gamma \ ,
\end{equation}  
leading to an observable final state comprising of a monophoton with 
missing energy-momentum. An obvious background to this is given by 
\begin{equation}
 e^+ e^- \to  \sum_{i} \nu_i\bar{\nu_i} + \gamma \ .
\label{eqn:bkgd_nu}
\end{equation}
In addition, final states where one has missed a putative visible
particle can also contribute. The leading such processes
are\footnote{It is easy to see that final states such as $f \bar
    f \gamma$, where $f \neq e$ is an arbitrary charged fermion, are
    rather subdominant.}
\begin{equation}
\label{eqn:bkgd_others}
\barr{rclcl}
 e^+ e^- &\rightarrow& \slashed{\gamma} + \gamma & \qquad & 
     \text{with one photon missing,} \\[1ex] 
 e^+ e^- &\rightarrow& \slashed{\gamma} + \slashed{\gamma} + \gamma, 
& & \text{where two of the photons are missing}, 
\\[1ex]
 e^+ e^- &\rightarrow& \slashed{e^+} \slashed{e^-} + \gamma, 
& & \text{where the leptons are missing}. 
\earr
\end{equation}
In the above ``missing'' implies that at least one of three conditions
hold, namely {\em (a)} the energy of the said particle is below the
threshold energy of the detector; {\em (b)} it travels along a path
lying outside the angular coverage of the detector; {\em (c)} it is
too close to another particle to be resolved, as a separate entity, by
the detector or {\em (d)} the particles is lost in the gaps between
the electromagnetic calorimeter (ECL) segments and/or an endcap. Given
the simple final state we are looking for, the third possibility is
very rare indeed. Similarly, with a substantial cut on the missing
energy-momentum, for {\em (a)} above to contribute would require the
emission of multiple such particles, and has only a small
probability. The dominant backgrounds, thus, are those in which the
event contains a second, and, maybe, a third (or more) photon, which
either fall(s) outside the angular coverage, or, more importantly,
fall(s) within the detector but go(es) undetected into the gaps~\cite{Kou:2018nap}.
Hence, the total background
composition is strongly dependent on the details of the
detector geometry. However, inspiration may be drawn
from the search for a Dark Photon~\cite{Kou:2018nap}, wherein the
major backgrounds for this search were found to arise from high cross
section QED processes such as $e^+e^-\to e^+e^-\gamma(\gamma)$ and
$e^+e^-\to\gamma\gamma(\gamma)$ with all particles, except for a
single photon, going undetected. In adopting the strategy and the
background rates from ref.~\cite{Kou:2018nap}, care must be taken,
though, to account for the fact that the study in
ref.~\cite{Kou:2018nap} was based on the dataset corresponding to an
integrated luminosity of only 20\,fb$^{-1}$. As it would turn out, our
simulations lead to a noticeably larger background count. Indeed, to
suppress the backgrounds to their levels, we need to impose cuts
stronger than they have done.  Nonetheless, a comparison with
ref.~\cite{Kou:2018nap} constitutes an useful countercheck and we
incorporate this in our study.

We begin by briefly recounting the details of the experimental
setup. The KEKB-II accelerator system collides a beam of $e^+$ with
an energy of 4 GeV against an electron beam of energy
7 GeV. We consider the direction of the latter as the reference
against which the polar angle is measured.  For the Belle-II detector,
we have
  \begin{itemize}
    \item \textbf{ECL coverage}: The electromagnetic calorimeter has
      an angular coverage of $(12.4^\circ, 155.1^\circ)$. In other
      words, $e^\pm, \gamma$ closer to the beampipe (in either
      direction) would not be registered. 
   \item \textbf{ECL gaps}: In addition, the ECL has gaps
     between the endcaps and the barrel at polar angle ranges
     $(31.3^\circ, 32.2^\circ)$ and $(128.7^\circ,
     130.7^\circ)$. Associated with extremely low detection
     efficiencies, particles falling in these gaps would not be
     registered and would essentially contribute to missing momentum.
   \item \textbf{Energy threshold}: For a $e^\pm, \gamma$ falling within 
     the `live' part of the ECL to be visible, it
     should have energy greater than $0.2$ GeV.
   \item \textbf{Trigger}: Furthermore, for a event to be triggered,
     at least 2 GeV of particulate energy needs to be deposited in the
     $(18^\circ, 140^\circ)$ window (other than in the dead zone). The corresponding trigger efficiency is 95\%.
  \end{itemize}

No detector, of course, has infinite resolution. For the Belle-II detector,
the energy resolution of the ECL is given by~\cite{Abe:2010gxa}
\begin{equation}
\frac{\sigma_E}{E} = \frac{0.066 \%}{\sqrt{\tilde E}}
   \oplus \frac{0.81 \% }{\tilde E^{1/4}}
   \oplus 1.34\%  \ ,
\qquad \tilde E \equiv \frac{E}{1 \gev} \ ,
\label{eqn:pgsparam1}
\end{equation}
where the different components are to be added in quadrature and
$\sigma_E$ represents a Gaussian smearing.  The relative angular
resolution is much finer, and is of little concern to us, as it
  contributes but little to the mismeasurement of momentum.

  \subsection{1D normalised distributions}
\begin{figure}[htb]
  \centering
  \begin{tabular}{cc}
  \includegraphics[width=0.42\textwidth,clip]{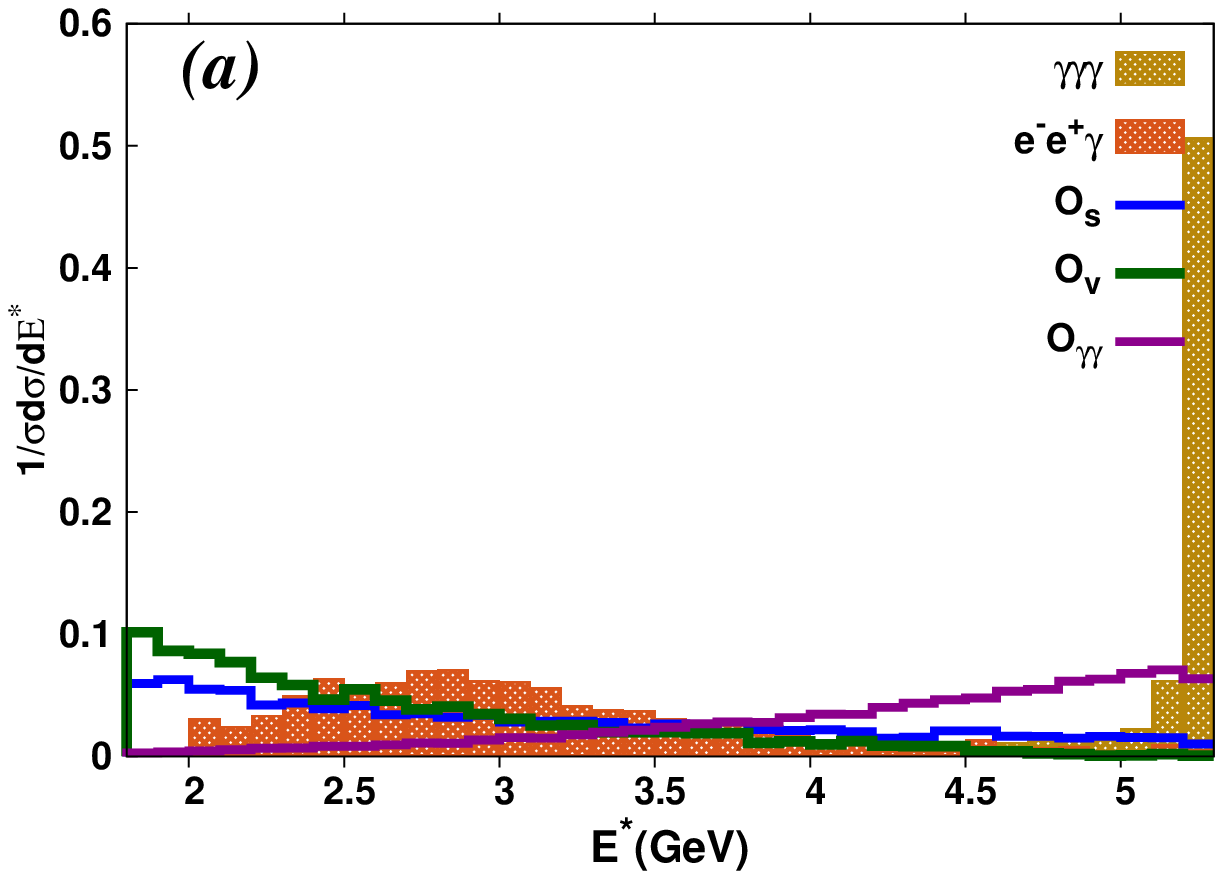} &
  \includegraphics[width=0.42\textwidth,clip]{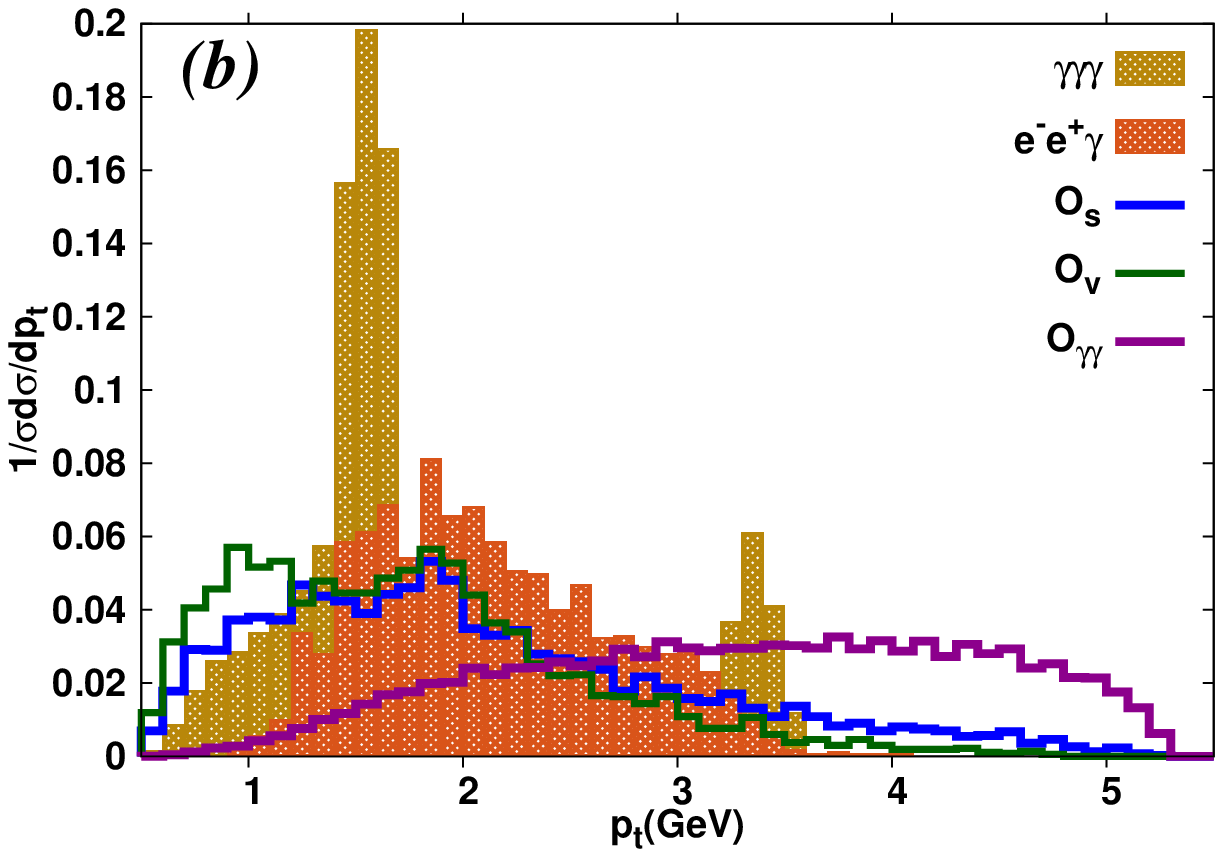} \\
  \includegraphics[width=0.42\textwidth,clip]{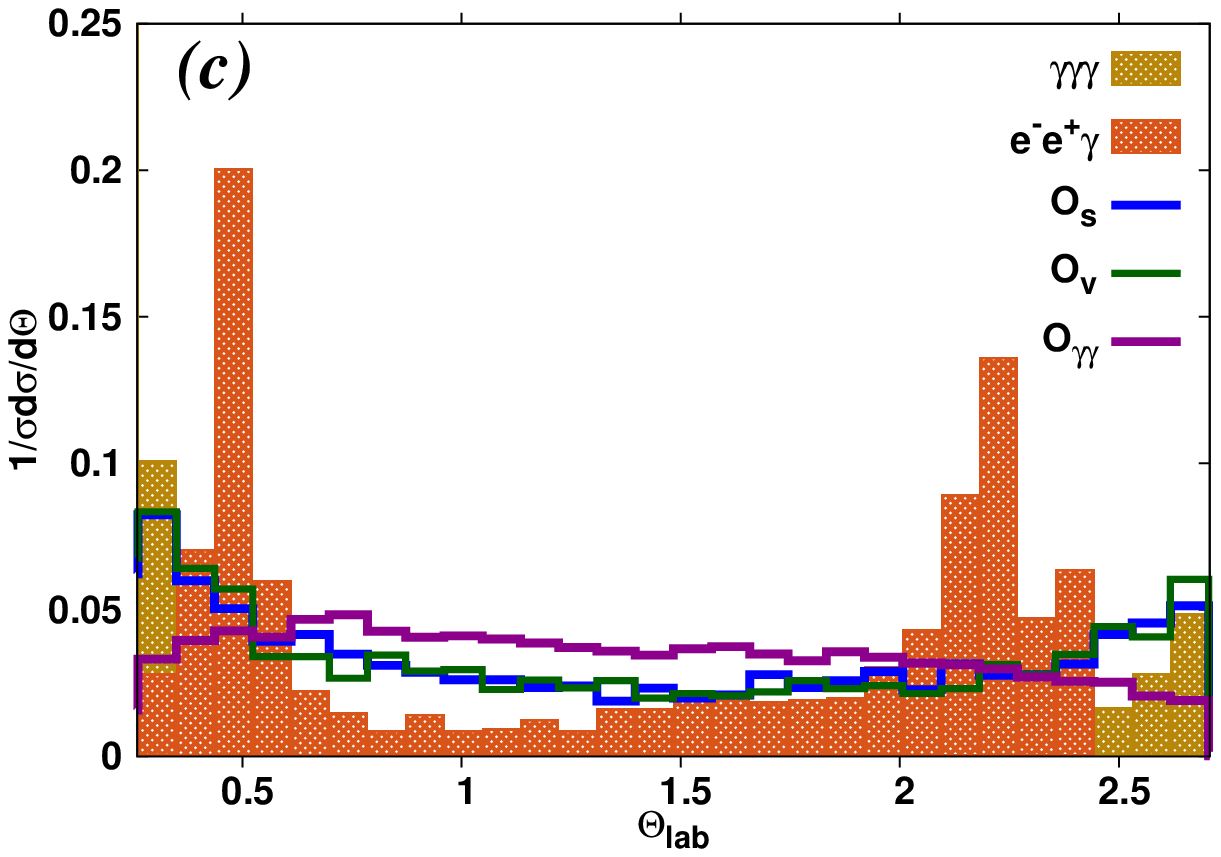} &
  \includegraphics[width=0.52\textwidth,clip]{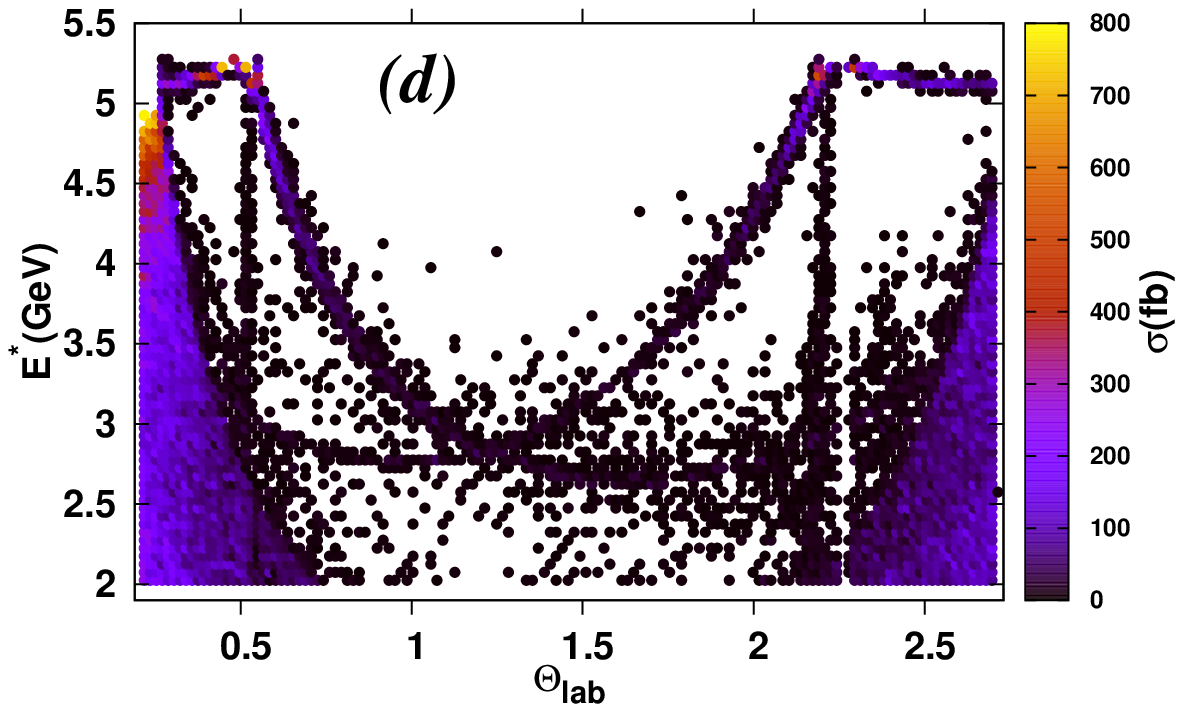}
  \end{tabular} 
  \caption{\small \em{\em (a-c)} Normalized 1-D differential distributions of
      kinematic observables $E^*$, $p_T$ and $\theta_{\rm lab}$
       for the leading photon (highest $p_T$)
      corresponding to the various background
      processes(without initial and final state radiation) obtained
      after applying basic cuts. Also shown is the signal
      process for the different operators, each corresponding to
       $m_{\varphi}=$100 MeV and {\em (d)} 2-D differential distribution for total 
       background. }
\label{fig:distribution} 
\end{figure}

Before we proceed further, let us examine the phase space
distributions (for signal and background) for the leading visible
photon in the final state, irrespective of the detector geometry
details.  In Fig.\ref{fig:distribution}, we display the normalized
distributions in the center-of-mass energy $E^*$, the transverse
momentum $p_T$ and the scattering angle $\theta_{\rm lab}$ for both
the signal (for a particular $m_\varphi$) and the background. Also
presented is a scatter-plot corresponding to a particular double
differential distribution. To facilitate an easier appreciation,
  we have deliberately switched off the initial and final state
  radiations as far as this figure is concerned.

Let us try to understand the distributions. To begin with, consider
$e^+ e^- \to \gamma\gamma$, where the photons, perforce, would have
a center-of-mass energy of $E^* =
\sqrt{s}/2$.  This particular background (already small as the second
photon would be missed in only a small fraction of events) can, of
course, be trivially eliminated by vetoing photons with $E^*$ close to
$\sqrt{s}/2$.  As for the $e^+ e^- \to 3 \gamma$ background, this can
be thought of as an additional photon being radiated off in the basic
$e^+ e^- \to \gamma\gamma$ process. This immediately tells us why this
background still peaks close to $E^* =
\sqrt{s}/2$ for the leading photon (as is seen in
Fig.\ref{fig:distribution}).  The $e^+ e^- \to e^+ e^- \gamma$
process, on the other hand, essentially consists of a photon radiated
off in a Bhabha scattering and, hence, is dominated by relatively
low-energy photons. As for the signal events, note that only for
${\cal O}_{\gamma\gamma}$ does the cross section increase with
$E^*$. This is easy to understand as the very structure of the matrix
element mandates this growth, at least in the absence of cuts.  As for
angular distribution of the backgrounds, clearly, in the center of
mass frame, it would be highly-peaked in both the forward and the
backward directions. That this is more peaked in the forward ($e^-$)
direction, is an obvious consequence of the larger energy, in the
laboratory frame, of the $e^-$ beam.

As we have already argued, the background processes are dominated by
amplitudes where the photon leg(s) is(are)
associated with soft and collinear singularities. Similar is the case
for the signal processes corresponding to the fermionic operators
({\it i.e.}, the first four) in eq.(\ref{the_operators}). Thus, the
event-distributions for all these cases would be dominated by
final-state configurations with low-$p_T$ photons. For the last two
operators in eq.(\ref{the_operators}), though, the DM particles come
off the photon and, hence, the latter must be imbued with a
non-negligible $p_T$.  The consequent distribution is quite
distinctive and is given by
\[
 \, \frac{d^2\sigma}{dp_Tdm^2_{\varphi\varphi}} =  \frac{\alpha_{\rm em}}{\pi^2\Lambda^4}\frac{p_T}{s} \,
              \left(1-\frac{4m^2}{m^2_{\varphi\varphi}}\right)^{1/2} \,\frac{\left[ 
              \left(1-\frac{m^2_{\varphi\varphi}}{s}\right)^2-\frac{2p_T^2}{s}\right]}{\left[ 
              \left(1-\frac{m^2_{\varphi\varphi}}{s}\right)^2-\frac{4p_T^2}{s}\right]^{1/2}}
\]
where $m^2_{\varphi\varphi}$ is the 
invariant mass for the (invisible) scalar-pair and is uniquely related to $E_\gamma$. For $s\gg 4 m^2$, this can be trivially integrated to yield
\[
 \sigma^{-1}\, \frac{d\sigma}{dp_T} =  12 \, \frac{p_T}{s} \,
              \sqrt{1-\frac{4p_T^2}{s}} \ .
\]
This feature could, in principle, be used not only to enhance
  the signal-to-noise ratio, but also, in the event of discovery, to
  distinguish between the fermionic and the photonic operators.

Understandably, the normalized profile for the signal ($e^+
e^-\rightarrow \varphi^* \varphi + \gamma$) events would look
remarkably similar to that for the last-mentioned background,
especially for very light DM. With increasing $m_\varphi$ though,
differences emerge, most notably for the photon-spectrum endpoint.
We show the distributions only for
$\mathcal{O}^f_s$, $\mathcal{O}^f_v$ and $\mathcal{O}_\gamma$ as
the ones for $\mathcal{O}^f_p$, $\mathcal{O}^f_a$ and
$\mathcal{O}_{\tilde{\gamma}}$ are, respectively,
almost identical to those for the former
operators. 

Finally, in Fig.\ref{fig:distribution}{\em (d)}, we present a
background event scatter plot in the plane spanned by $E^*$ and
$\theta_{\rm lab}$. As we have already discussed, the background
processes are dominantly concentrated in parts of the phase space
where the photon either has low energy or is travelling reasonably
close to the beam pipe.  With the background being demonstrably small
in the region $30^\circ <\theta_{\rm lab} < 130^\circ$, we use this as
a selection cut, as has been advocated in the Belle physics
book~\cite{Kou:2018nap}.

At this stage, we would like to point out that our simulations
  of the backgrounds, with ostensibly the same kinematic cuts, leads
  to a larger cross section than that presented in
  ref.\cite{Kou:2018nap}. For example, as a comparison of
  Fig.\ref{fig:distribution}$(d)$ with Fig.204 of the said reference
  shows, the latter is almost totally bereft of the dense curved arm
  in the region $\pi/2 \lapp \theta_{\rm lab} \lapp 9/4$. Presumably,
  such events were excluded on the back of detailed detector-level
  simulations that have not been spelled out. In the absence of such
  information, we must accept the larger backgrounds as represented by 
Fig.\ref{fig:distribution}$(d)$. We would, subsequently, seek to suppress 
this by the imposition of a strong cut on the photon transverse momentum,
one that is absent in ref.\cite{Kou:2018nap}. This, however, would 
eliminate a non-negligible fraction of the signal events as well.

\subsection{Selection Cuts}
  Based on the 1D distributions, we impose only a simple set of {\bf
    selection cuts}.  An event should contain one and only one photon
  satisfying
   \begin{equation}
   p_t > 3.5 \gev \ , \qquad 
   30^{\circ} \leq\theta_{\gamma}\leq 130^{\circ} \ .
   \end{equation}
These cuts are chosen to reject the background due to {\em (a)} the
three photon final state, where two of the photons are not registered;
typically this is dominated by the case where one of the photons makes
a relatively small angle with the beam pipe while the second could
make a slightly larger angle but still fall outside the ECL coverage
area, or hit one of the ECL gaps and {\em (b)} analogous radiative
Bhabha events with both $e^\pm$ being missed similarly. With these
cuts, we study the production of scalar DM masses up to
$m_{\varphi}=1$\,GeV. In this work, we have generated the
signal Monte Carlo \hbox{\texttt{MadGraph5\,v2.2.2}}
\cite{Alwall:2014hca} and the background events using Babayaga
\cite{Balossini:2006wc,CarloniCalame:2000pz,CarloniCalame:2003yt,Balossini:2008xr}.

\subsection{The analysis}
\begin{figure*}[tb]
   \centering
   \begin{tabular}{cc}
   \includegraphics[width=0.45\textwidth,clip]{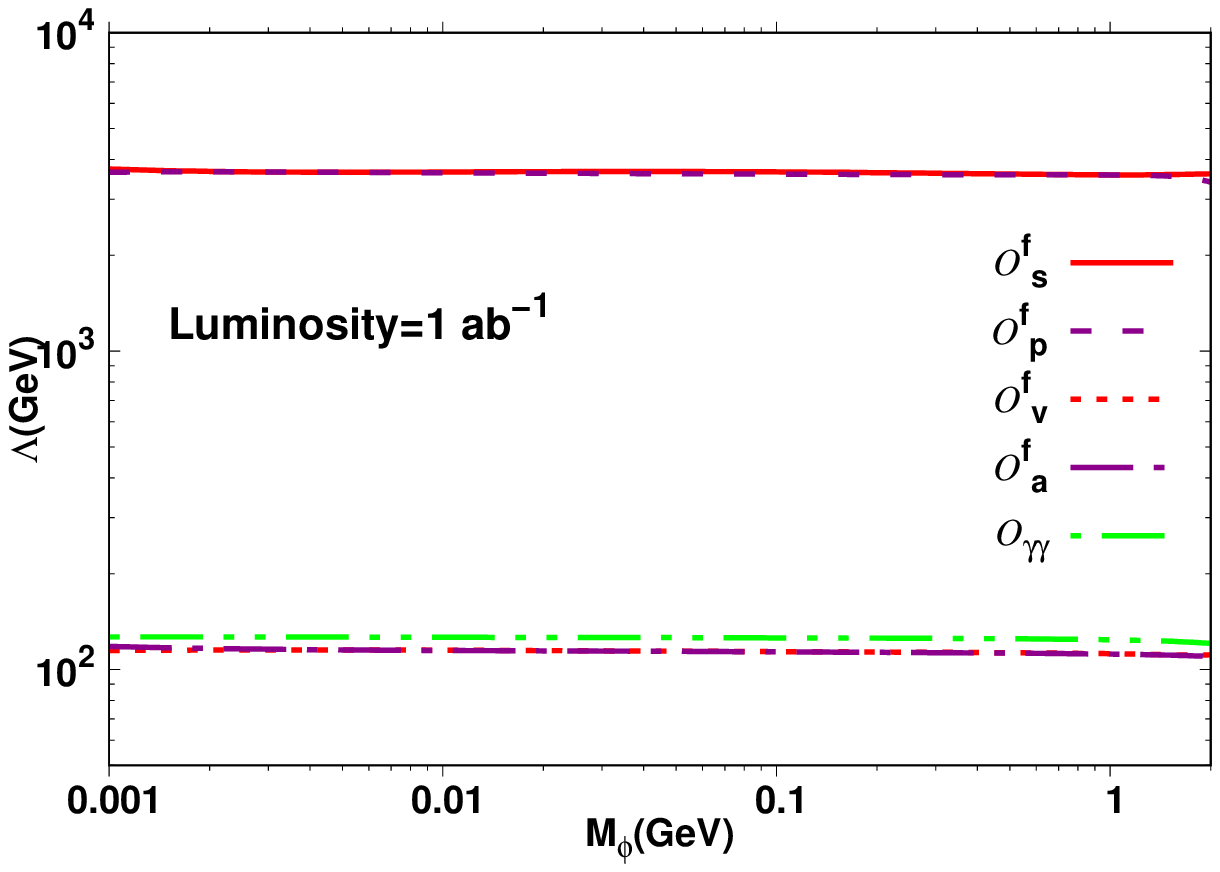} &
   \includegraphics[width=0.45\textwidth,clip]{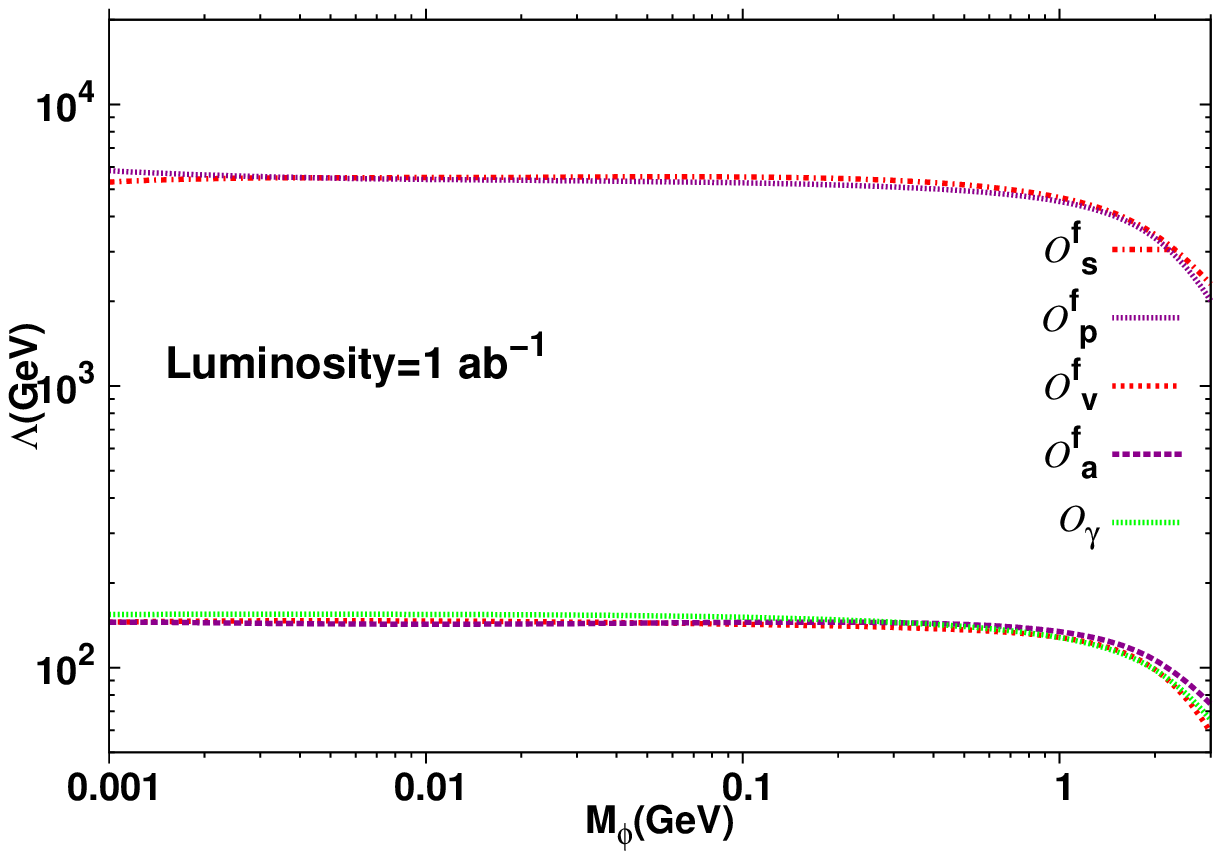} \\
   \includegraphics[width=0.45\textwidth,clip]{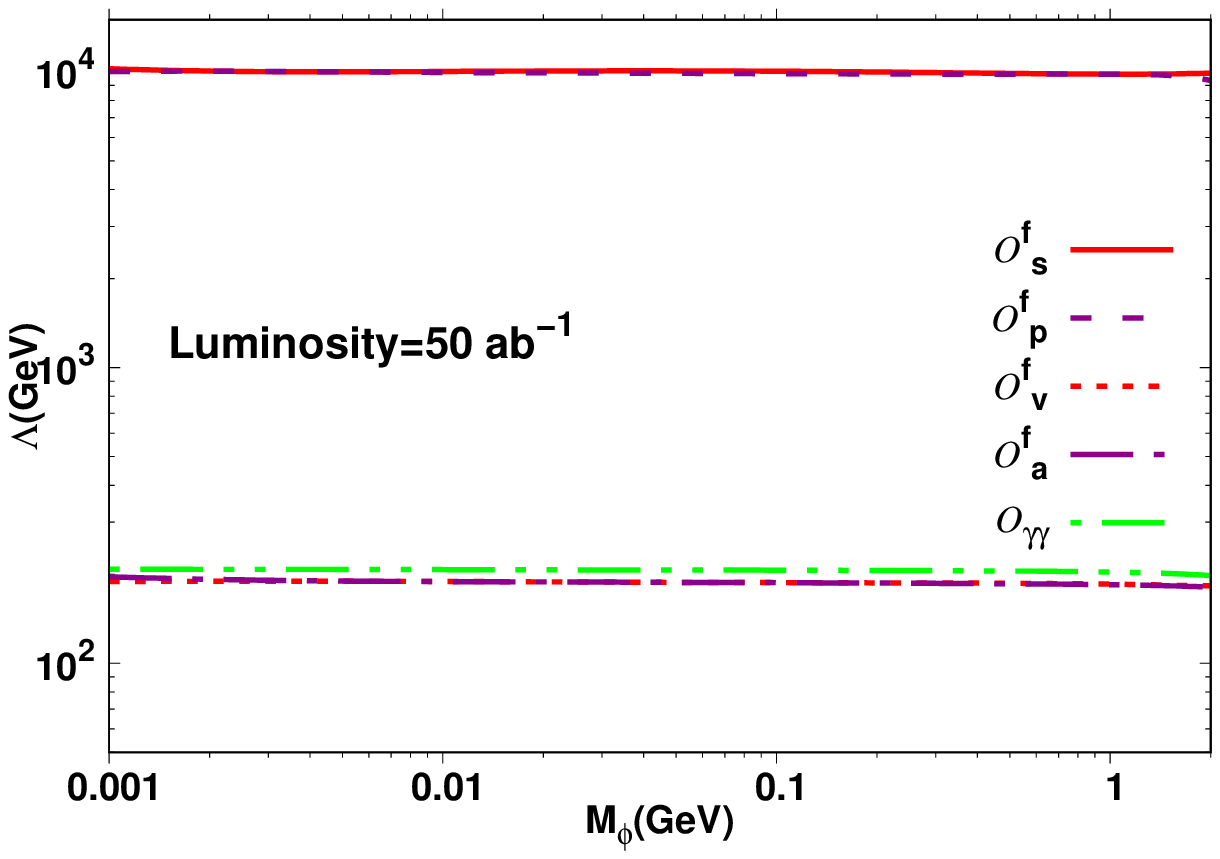} &
   \includegraphics[width=0.45\textwidth,clip]{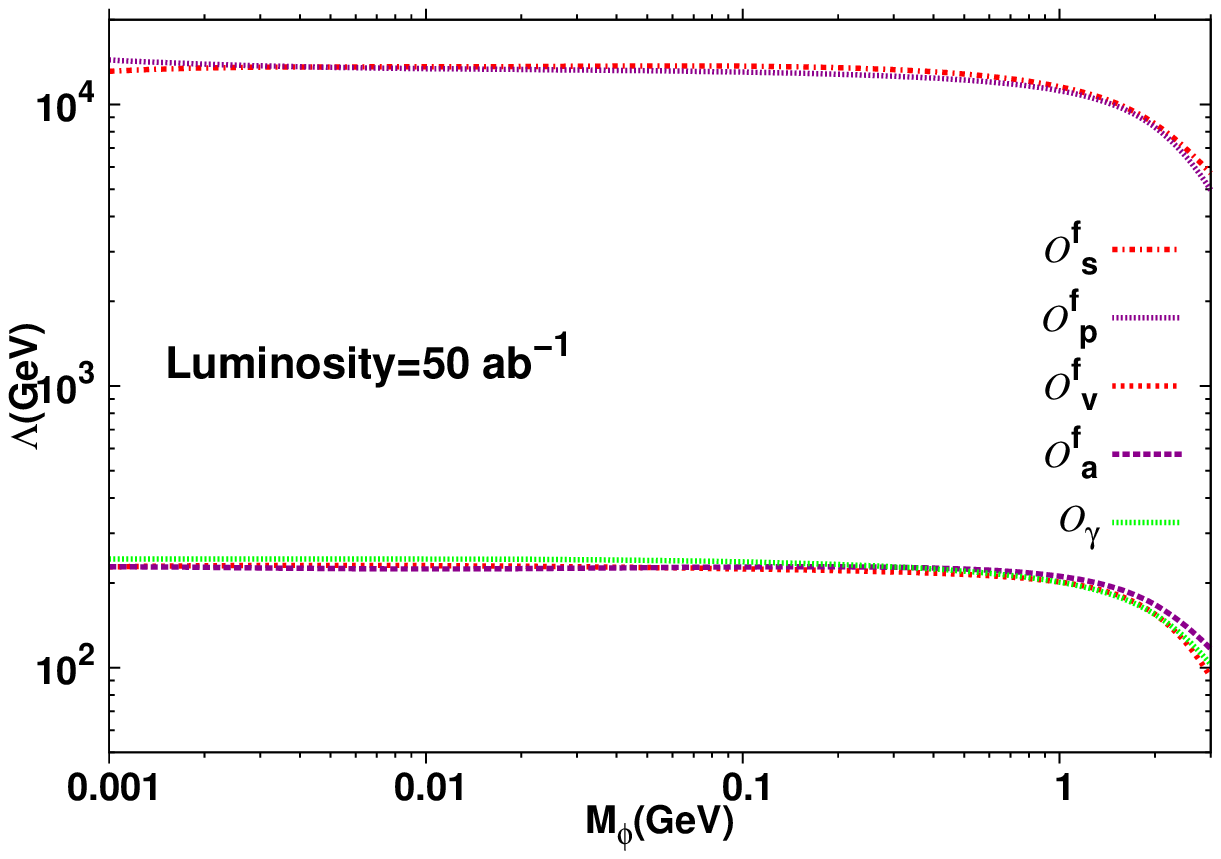} 
   \end{tabular}
   \caption{\small \em 99.7\% C.L. contours in the $m_{\varphi}$---$\Lambda$
       plane. The left panels are obtained using the $\chi^2$
       analysis. The right panels are obtained using simple
       $S/\sqrt{B}$ criteria, but with the much smaller backgrounds of
       Ref.\cite{Kou:2018nap}. The upper(lower) panels correspond to
       integrated luminosity of $1 \, (50) \, {\rm ab}^{-1}$.}
 \label{fig:contour} 
 \end{figure*}

Imposing the aforementioned cuts, we now perform a
$\chi^2$-test, with the statistic defined as
\begin{equation}
\chi^2 =
\sum_{i=1}\sum_{j=1}
\left(\frac{ N_{ij}^{\rm NP}}{N_{ij}^{\rm tot}}\right)^2\ .
\end{equation}
Here, $ij$ denotes a particular bin with $N_{ij}^{\rm NP}$
($N_{ij}^{\rm tot}$) being the number of signal events (total number
of events) in the bin. With the photon being the only visible
particle, we have only two independent phase space variables
associated with the final state.  In particular, we choose to work
with the two-dimensional distribution defined by the center-of-mass
energy $E^*$ and the laboratory-frame scattering angle $\theta_{\rm
  lab}$, and divide the associated space into
  uniform bins of size $0.1\times 0.2 \gev$ each. With
the SuperKEKB slated to deliver a peak luminosity of $\sim 8 \times
10^{35} \, {\rm cm}^{-2} \, {\rm s}^{-1}$ (or, an integrated
luminosity of $\sim 8 \, {\rm ab}^{-1}$ per year, for a nominal year
of $10^7\,$s)~\cite{Abe:2010gxa}, we consider a
representative\footnote{Clearly, this choice is conservative and
  correspond to the initial phase.} value of the total integrated
luminosity, namely $\mathcal{L} $ = 1 {\rm ab}$^{-1}$, allowing us to
obtain the corresponding reach/sensitivity of the experiment.  This is
displayed, in the form of $3\,\sigma$ contours in the
$m_{\varphi}-\Lambda$ plane, in Fig.~\ref{fig:contour}. For
comparison, we also present the reach obtainable using a simple
$S/\sqrt{B}$ statistic but using the considerably smaller background
estimates of ref.\cite{Kou:2018nap}. It is intriguing to see that the
two sets of contours differ by only about 25\%.

\subsection{Discussion}
That the sensitivity to operators ${\cal O}^f_{s,p}$ are much higher
than that to the rest is but a reflection of the fact that the former
are only dimension-five, while the rest are dimension-six. Similarly,
the insensitivity to the parity structure (scalar versus
pseudoscalar and vector versus axial vector) can be understood by
realizing that, with $\sqrt{s} \gg m_e$, these differences between the
interactions would only have been manifested had we considered
polarized beams. Indeed, were a signal to be established, the use of
polarization would be invaluable in the unravelling of the underlying
interaction.

The large sensitivity to operators ${\cal O}^f_{s,p}$ renders this
experiment one of the best for such small DM masses. The fall-off for
$m_\varphi \gapp 3 \gev$ is, of course, expected on kinematic grounds.
And while the sensitivities to operators
${\cal O}_{v,a,\gamma,\tilde{\gamma}}$ are not as high, they are
still better than those achievable
at other current collider experiments. Although, naively, it could be
argued that new physics at $\Lambda \sim 250 \gev$ should have been
visible at, say the LHC, this is not necessarily true if the DM were
hadrophobic. At a future high-energy linear collider, however, even
such a scenario should be manifestly visible.

\section{Monophoton searches at LEP.}
Temporarily turning to other colliders, both
present and past, we consider, next, the
potential of each in this context. At the LHC, such a light DM would
manifest itself essentially just as neutrinos
do, but with considerably smaller cross-sections (which would be
further suppressed if the coupling of $\varphi$ to the
first-generation quarks are sub dominant). Given the fact that, at a hadronic collider as
complicated as the LHC, there are numerous other sources for missing
transverse momentum, the sensitivity is low indeed. Much the same was
true for the erstwhile Tevatron too.

At the LEP, though, the prospects were much better, courtesy the
extremely clean environment. Indeed, neutrino number-counting was one
of the successes of the four experiments. The most sensitive test was
the lineshape at the $Z$-peak, and led to $N_\nu = 2.9840 \pm
0.0082$~\cite{ALEPH:2005ab}.  This deficit, nominally, would impose a
very strong bound on any extra sources of missing
energy-momentum. However, this clearly is of little consequence in the
present context as the DM would not really manifest itself at the
$Z$-peak and, hence, in the lineshape.  On the other hand, monophoton
searches (exactly analogous to what we propose at
Belle-II) were sensitive indeed. In fact, such a search was also
performed {\em just below} the $Z$-peak (primarily, when the collider
was being ramped up) and constituted the first worthwhile neutrino
number-counting exercise leading to $\delta N_\nu \sim 0.1$.

Much better bounds are available, though, from dedicated mono-photon
searches at LEP-II, where one looked for highly-energetic photons in
association with missing energy resulting from the process $e^+ e^-
\rightarrow \gamma + (\text{invisible})$. The major background to this
process is, obviously, $e^+ e^- \rightarrow \nu_i + \bar{\nu_i}
+\gamma$ and, unlike in the case of Belle-II, dominates the
instrumental background.  In particular, we draw inspiration from a
particular study at DELPHI~\cite{Abdallah:2008aa} based on an
integrated luminosity of $650\,\mathrm{pb^{-1}}$ at $\sqrt{s}$ between
180 GeV and 209 GeV.  To determine the sensitivity, we have executed
an analysis similar to that in ref.~\cite{Fox:2011fx}, implementing
detector efficiencies and resolution as detailed in
ref.~\cite{Abdallah:2003np}. We, however, effect one simplification
(one that facilitates both presentation and an easy
understanding). Rather than simulate events for each of the
actually implemented $\sqrt{s}$ values in the 180--209 GeV
range, we consider the weighted (with the respective integrated
luminosities) mean and simulate events only for $\sqrt{s} =
200\gev$. This reproduces very well the SM background as obtained in
Ref.\cite{Abdallah:2008aa}, except very close to the edge of the phase
space, {\em viz.} $x_\gamma\equiv E_\gamma/E_{\rm beam} \gapp
0.9$. The remaining small discrepancy is as much a consequence of our
inability to effect a full detector simulation as that of the
aforementioned approximation. To avoid such effects, we shall omit
events with $x_\gamma > 0.9$ from our analysis.  For our simulations
of the SM background as well as the signal, we use
MadGraph5~\cite{Alwall:2011uj} in conjunction with an appropriate
implementation of FeynRules~\cite{Alloul:2013bka}. Having standardised
this (by comparing with Ref.\cite{Abdallah:2008aa}), we use the latter
({\em i.e.}, the DELPHI Monte Carlo) for the background events,
thereby ensuring a very accurate rendition of the same.

Before we decide on phase space cuts {\it etc.}, we must decide on the
triggers. At DELPHI, three different triggers were used to select
single-photon events.  Events with a photon with polar angle in the
range $45^\circ < \theta < 135^\circ$ were detected in the High Density
Projection Chamber (HPC) with an energy threshold of
$E_{\gamma} = 6 \gev$. The trigger efficiency for photons in the HPC,
in the analysis, was assumed to increase linearly from 52\% at $
E_{\gamma} = 6 \gev$ to 77\% at 30 GeV, and then to 84\% at 100
GeV. This trigger efficiency is then multiplied by the reconstruction
and analysis efficiency, which was assumed to increase linearly from
41\% at 6 GeV to 78\% for $E_{\gamma}= 80 \gev$ and constant
thereafter.

The Forward Electromagnetic Calorimeter (FEMC), located at $12^\circ <
\theta < 32^\circ$ and at $148^\circ < \theta < 168^\circ$ 
could accept events with a single photon of energy $E_{\gamma}> 10\gev$.  The
corresponding trigger efficiency increases approximately linearly from
93\% at 10 GeV to 100\% at 15 GeV and above, and then it is multiplied
by the analysis efficiency (related to reconstruction and event
selection efficiency) which increases linearly from 57\% at 10 GeV to
75\% at 100 GeV. This has to be further multiplied by 89\% to account
for the additional loss of events due to noise and machine background.
Very forward ($3.8^\circ < \theta < 8^\circ$ or $172^\circ <
  \theta < 176.2^\circ$) photons with an energy threshold of
30 GeV produced a signal in the Small Angle Tile
Calorimeter (STIC).  Here, the efficiency is assumed to be
48\%, based on ref.~\cite{Fox:2011fx}. This is then multiplied by
overall analysis efficiency of 48\%.

In estimating the measured energy from the simulated one, we need to
incorporate the resolution of the electromagnetic calorimeter, and
this was given by~\cite{Abdallah:2003np}
\begin{equation}
\barr{rclcl}
\dis \frac{\sigma_E}{E}(HPC) &=& \dis 0.043 \oplus \frac{0.32}{\sqrt{E}}
	\\[2.5ex]
\dis \frac{\sigma_E}{E}(FEMC) &=& \dis 
0.03 \oplus \frac{0.12}{\sqrt{E}} \oplus \frac{0.11}{E}	
\\[2.5ex]
\dis \frac{\sigma_E}{E}(STIC) &=& \dis 0.0152 \oplus \frac{0.135}{\sqrt{E}}
\earr
\end{equation}
where the energy was measured in GeVs and the three
contributions added in quadrature. The errors due to
the finite angular resolution were too small to be
of any consequence.

\begin{figure}[htb]
\centering
\includegraphics[width=95mm]{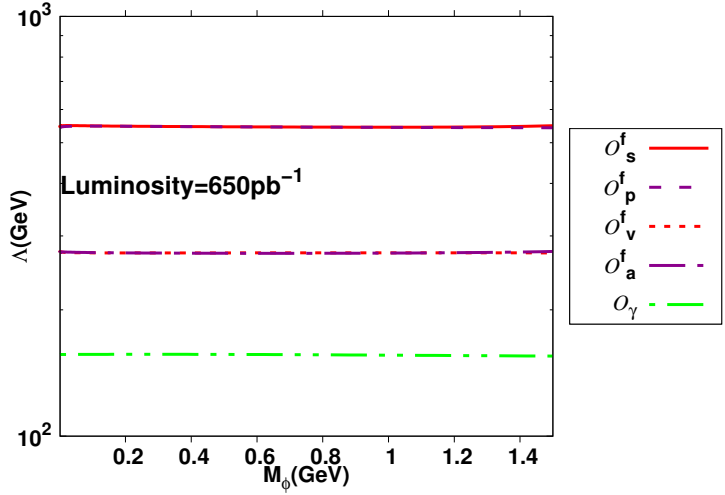}
  \caption{\small \em{99\% Confidence Level contours on the $m_{\varphi}$ and
$\Lambda$ plane from the $\chi^2$ analyses with an integrated luminosity of 
 $650 \,{\rm pb}^{-1}$.}}
\label{fig:contour_LEP}
\end{figure}

Events with $x_\gamma < 0.06$ fell below the trigger threshold and were not registered. On
  the other hand, in the region $0.7 \lapp x_\gamma \lapp 0.9$, the
  background is highly enhanced on account of the radiative return to
  the $Z$ and, hence, contributes little to sensitivity to new
  physics. And, as already explained earlier, we altogether omit the
  $x_\gamma \gapp 0.9$ window from our analysis. The rest of the phase we 
  divide into $x_\gamma$ bins of width 0.05 each and compare the simulation 
  for signal events with the background as in Ref.\cite{Abdallah:2008aa}. 
To this end, we effect a $\chi^2$-test, with the statistic defined as 
\begin{equation}
\chi^2 =
\sum_{i=1} \left(
\frac{N_{i}^{\rm NP}}
     {\Delta N_{i}^{\rm tot}}\right)^2 \ .
\end{equation}
Here, $i$ denotes a particular bin with $N_i^{\rm NP}$ ($N_{i}^{\rm
  tot}$) being the number of signal events (total number of events) in
the bin and  $\Delta N_{i}^{\rm tot}$ is largely dominated by 
the SM background as in Ref.\cite{Abdallah:2008aa}. The $\chi^2$, 
thus calculated, can be translated to $3\,\sigma$
contours in the $m_{\varphi}-\Lambda$ plane, as displayed in
Fig.~\ref{fig:contour_LEP}.

Understandably, there is little dependence on $m_\varphi$, far less than
that at Belle-II. This is but a reflection of the fact that, for
center-of-mass energies as large as that at the LEP, a DM mass in the
range we are interested in is virtually indistinguishable from
zero. For very analogous reasons, the chirality structure of the
fermionic current (or the difference between $\mathcal{O}_{\gamma}$
and $\mathcal{O}_{\tilde \gamma}$) is immaterial. Similar to the
Belle-II limits, the sensitivity to the operators
$\mathcal{O}^f_{s,p}$ is much higher than that for the other
operators. This, of course, is primarily due to the fact that the
former are only dimension-five operators, while the others are
dimension six. Amongst the latter, naively, one would have expected
that $\mathcal{O}^f_{v,a}$ would lead to lower sensitivity than the
photonic ones on account of the fact that the signal matrix element
has a structure similar to that for the background\footnote{Strictly
  speaking, the similarity holds only for the $\bar\nu_\mu \nu_\mu
  \gamma$ and $\bar\nu_\tau \nu_\tau \gamma$ components of the
  background. The $\bar\nu_e \nu_e \gamma$ final state receives
  additional contribution from the $W$-diagram, leading to much larger
  cross sections. However, this additional complication is not germane
  to the issue.}. On the other hand, processes due to
$\mathcal{O}_{\gamma}$ and $\mathcal{O}_{\tilde \gamma}$ suffer from
additional $s$-channel suppression leading to smaller cross sections
and, hence, lead to lower sensitivities. Overall, it is easy to see
that despite LEP having large $\sqrt{s}$, the limits obtained for
Belle-II are much stronger.  This is but a
consequence of the much higher luminosity at Belle-II as well as the
virtual absence of the neutrino background.

\section{Complementary Signals at Belle-II}
   \label{sec:complementary}
   Until now, we have considered only the monophoton final state, both
   at Belle-II and at LEP, primarily because it constitutes the
   simplest search strategy.  Given the high luminosities achievable
   at Belle-II, it is, however, worthwhile to consider more
   complicated final states. Even accounting for a suppression of the
   cross section, as well as experimental issues (such as analysis
   efficiencies), these might, yet, lead to additional and nontrivial
   sensitivity. We, now, consider three such cases, each involving the
   sighting of one or more charged lepton $\ell \, (= e, \mu)$
   alongwith missing energy-momentum.

In particular, we turn our attention to three distinct cases, namely,
\begin{itemize}
 \item \textbf{Case I}: 
\begin{equation}
 e^- + e^+ \rightarrow \varphi + \varphi^* + e^- + e^+ \ ,
\label{case_1}
\end{equation}
or, in other words, Bhabha scattering with a pair of DM particles
being radiated off one of the four legs.
 \item \textbf{Case II}: 
\begin{equation}
 e^- + e^+ \rightarrow \varphi + \varphi^* + \mu^- + \mu^+ 
\label{case_2}
\end{equation}
While analogous to Case I above, this is simpler, and would turn out 
to be more sensitive than the former.
 \item \textbf{Case III}: 
\begin{equation}
 e^- + e^+ \rightarrow \varphi + \varphi^* + \mu + \slashed{\mu}. 
\label{case_3}
\end{equation}
This is very similar to Case II above, with the exception that only a 
single muon should be visible.
\end{itemize}
While an analogue of Case III could be defined with a single $e^-$ (or
$e^+$) instead, it has a low sensitivity and, hence, we are omitting
it. It should also be realized that each of Cases II and III could as
well be defined with the tau lepton instead of the muon. Indeed, the
analysis would be very similar, except for the fact that
tau-identification and/or reconstruction would be associated with a
further loss in efficiency. Consequently, the results for the muonic
channel can be trivially extended to the tauonic ones at the cost of
inclusion of such efficiency factors.

The corresponding irreducible backgrounds arise
primarily\footnote{While we have, for the sake of completeness,
  considered $e^- e^+ \to \ell^- \ell^+ \nu_i \bar\nu_i$, proceeding
  through a virtual $Z$ or $W$'s, the corresponding cross sections
  are, understandably, orders of magnitude suppressed.}  from
\begin{equation}
\label{eqn:bkgd}
\barr{rclcl}
 e^+ e^- &\rightarrow& n \slashed{\gamma} +  \ell + \bar\ell & \qquad & 
     \text{with all the photons missing,} \\[1ex] 
 e^+ e^- &\rightarrow& \nu + \bar{\nu} + \ell + \bar\ell
\earr
\end{equation}
with the understanding that, for Case III, one of the two leptons should 
also go missing.  

For the photons and the $e^\pm$, the requirements for one to be seen
(or, equivalently, being missed) remains, of course, as in
Sec.\ref{sec:monophot_belle}. The muons, though, escape the ECL and
are caught, instead by the KLM ($K_L$ and muon detector), which
consists of an alternating sandwich of thick iron plates (which also
serve as the return yoke for the magnetic flux from the
superconducting solenoid) and active detector elements
(glass-electrode resistive plate chambers). The consequent
instrumental requirements are 
 \begin{equation}
 E_\mu > 0.2 \gev \ , \qquad
25^\circ \leq \theta_{\rm lab}(\mu) \leq 155^\circ \ .  
\end{equation}
Note, in particular, that unlike in the case of the ECL, there are no
gaps in the KLM, and, hence, muons cannot thus escape the detector
unlike either of $e^\pm, \gamma$.  We generated the SM process $e^+
e^- \rightarrow \gamma + \ell + \bar{\ell}$ with BabaYaga~\cite{Balossini:2006wc,CarloniCalame:2000pz,CarloniCalame:2003yt,Balossini:2008xr}
and signal events with MadGraph5~\cite{Alwall:2014hca} with the following basic
cuts:
\begin{itemize}
 \item Minimum total transverse momentum of charged lepton(single or both), taken to be 1.0 GeV,
 \item Minimum energy of the charged leptons is taken to be 0.5 GeV,
 \item  $12.4^{\circ}<\theta(e^{\pm})<155.1^{\circ}$, \qquad  $25^{\circ}<\theta(\mu^{\pm})<155^{\circ}$ 
 \item $\Delta R(\ell^+,\ell^-) > 0.4$.
\end{itemize}


\subsection{1D normalised distributions}
To decide the cut strategy, we discuss next the normalized
distribution for various individual kinematic observables.  For case I
and case II, these are given in Fig.\ref{dist:ee} and
Fig.\ref{dist:mm} respectively, where we have, for reasons of brevity,
displayed only the leading and next-to-leading background. We discuss
each case in turn.

\subsubsection{Case I : the $e^- e^+ + \slashed{E}_T$ final state}
\begin{itemize}

\item As we see in Figs.\ref{dist:ee}{\em (a,d)} respectively, for the
  background events, the energies of the outgoing $e^\mp$ in the
  center-of-mass frame are quite different, with the $e^-$ generally
  being harder. This lack of symmetry is also reflected by the
  distributions for the respective scattering angles (in the
  laboratory frame), namely Figs.\ref{dist:ee}{\em (b,e)} and, to a
  slightly smaller degree, by the transverse momenta, {\em i.e.},
  \ref{dist:ee}{\em (c,f)}.

An understanding of this is best achieved by considering the lost
photon.  Discounting, for the time being, the gaps in the calorimeter,
the photon can be lost only if it either goes down the beam pipe or
has too small a energy.  Since we require that the missing energy be
substantial, the latter alternative is ruled out (unless the missing
energy momentum is shared by multiple missed photons, a final state
with only a small production cross section). Recognizing that the ECL extends to smaller
angles in the forward ($e^-$) direction than in the latter, it is
immediately obvious that the bending of the electron would, typically,
be much smaller than that suffered by the positron
(Figs.\ref{dist:ee}{\em (b,e)}).  Combined with the larger laboratory
frame energy of the initial $e^-$ beam, it translates to a smaller
degradation, even in the center-of-mass frame, of the electron
energy. On the other hand, the larger (smaller) initial energies of
the $e^-$ and the $e^+$, convoluted with the smaller (larger)
scattering angles implies that the transverse momenta are less
dissimilar.

\item While the discussion above encapsulates the leading behaviour of
  the background that owes itself to the ``$t$-channel'' part of the
  underlying Bhabha-scattering (with photons having been radiated
  off), it does not explain all the features, in particular the
  secondary peaks. These, though, can
  be readily understood once one includes the
  ``$s$-channel-like'' diagrams.

\begin{figure}[H]
  \centering
  \begin{tabular}{ccc}  
  \hskip -05pt
\includegraphics[width=0.33\textwidth,height=0.33\textwidth,clip]{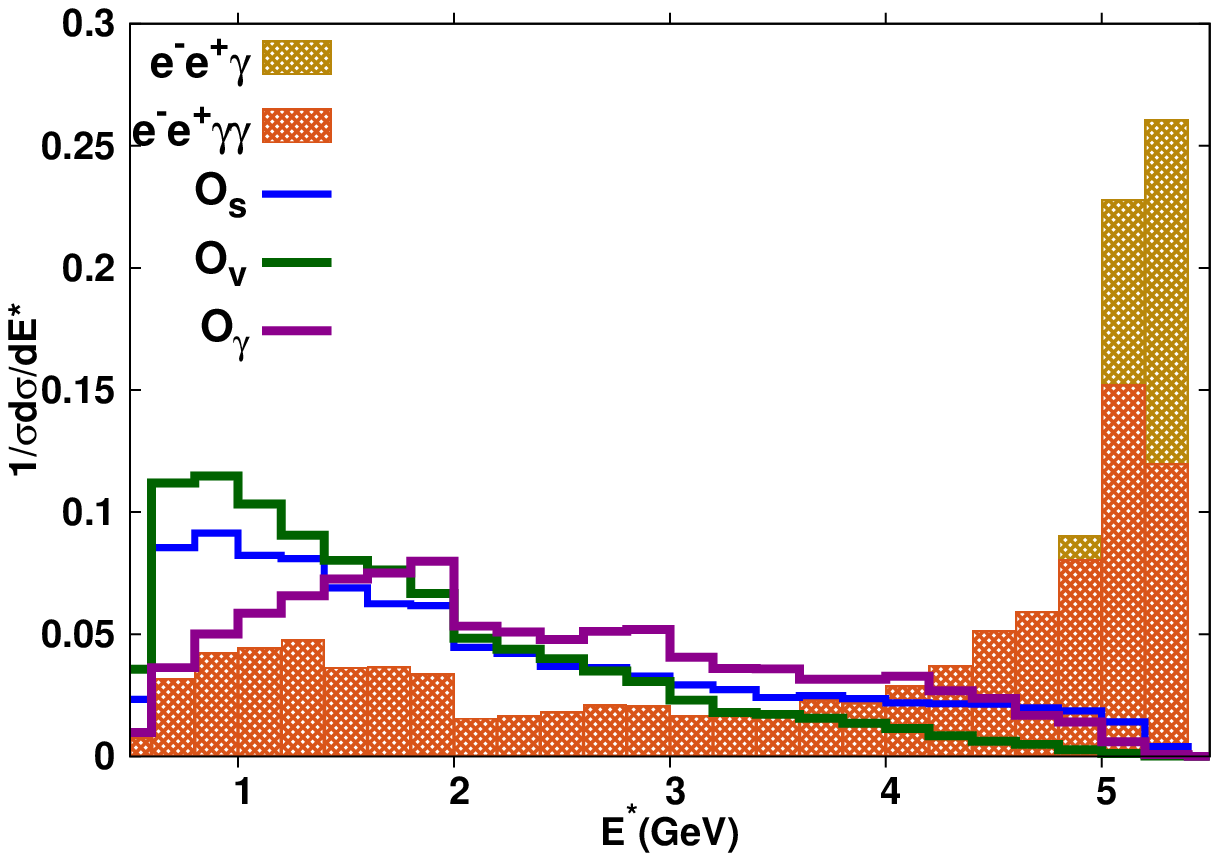} 
&  \hskip -05pt
  \includegraphics[width=0.33\textwidth,height=0.33\textwidth,clip]{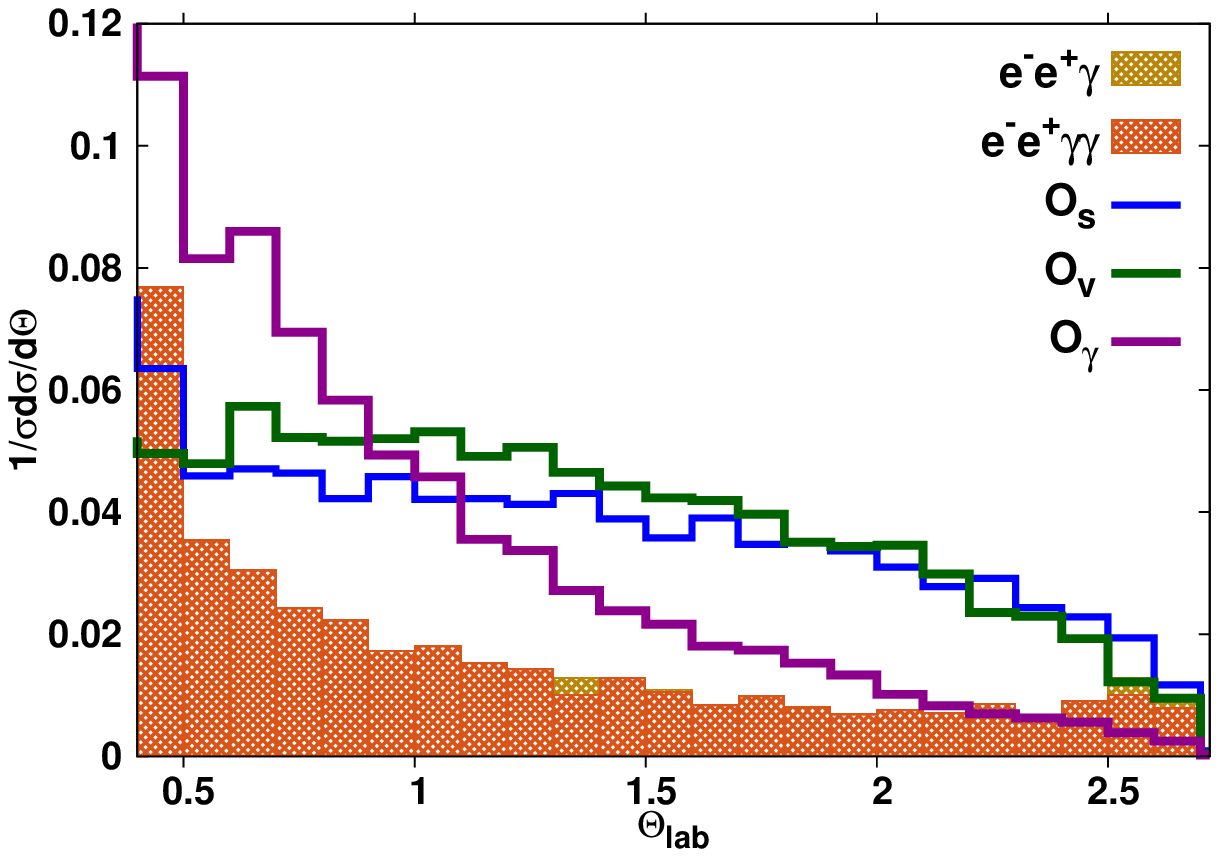} 
&   \hskip -05pt
  \includegraphics[width=0.33\textwidth,height=0.33\textwidth,clip]{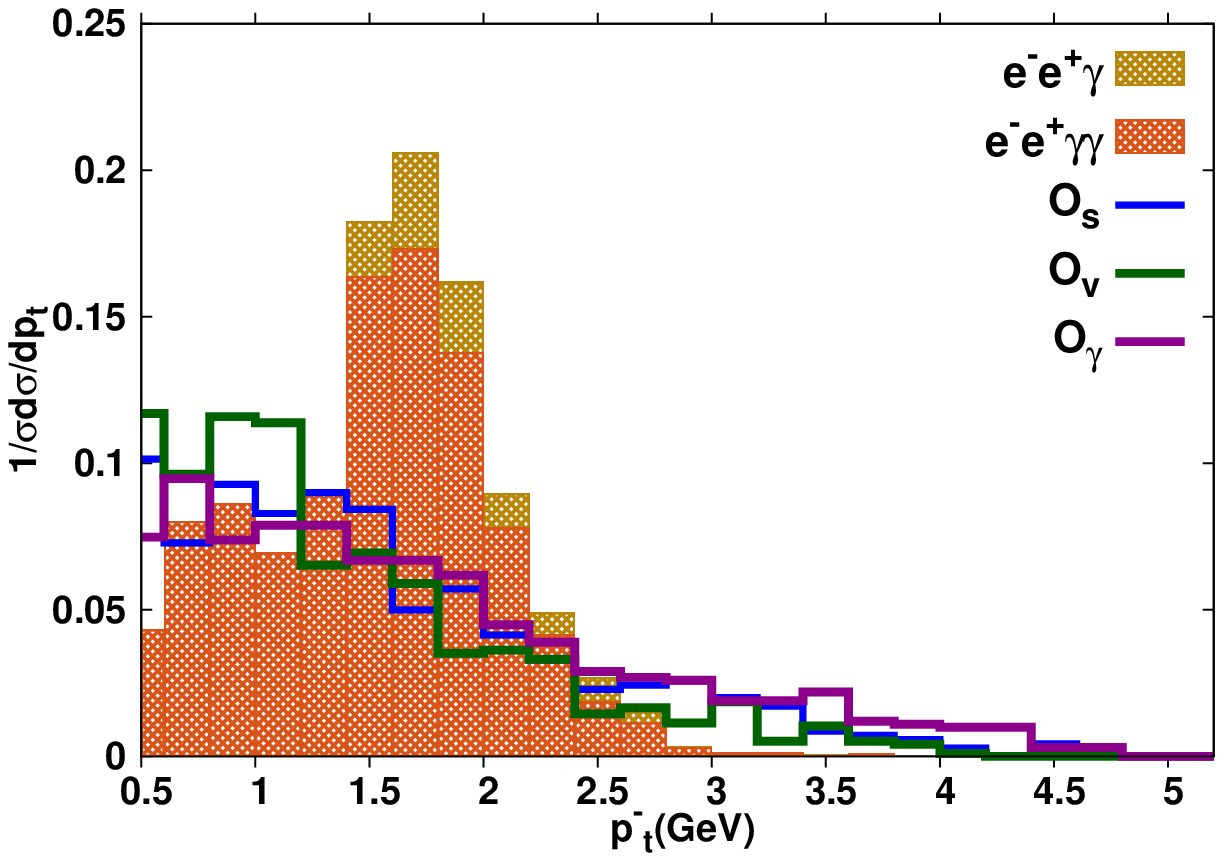} 
\\
  \hskip -05pt
  \includegraphics[width=0.33\textwidth,height=0.33\textwidth,clip]{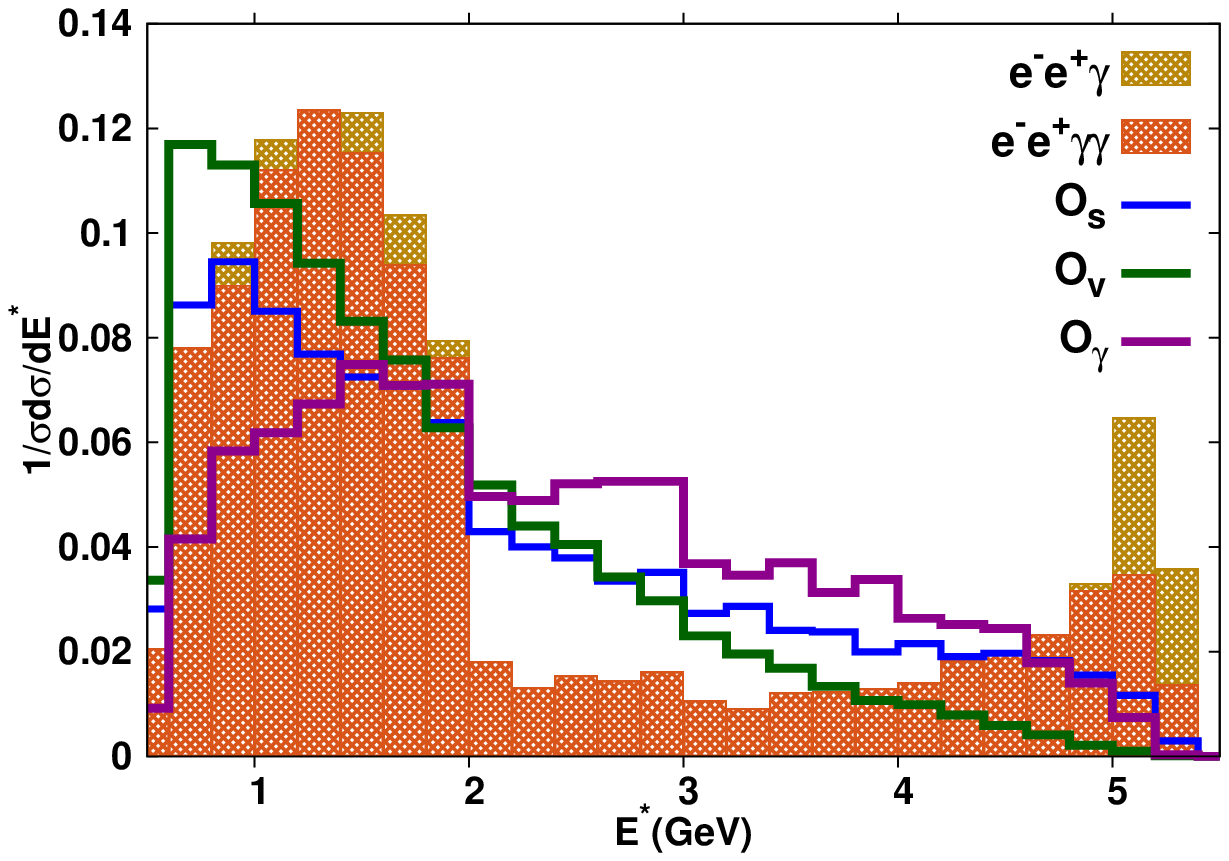} 
&  \hskip -05pt
  \includegraphics[width=0.33\textwidth,height=0.33\textwidth,clip]{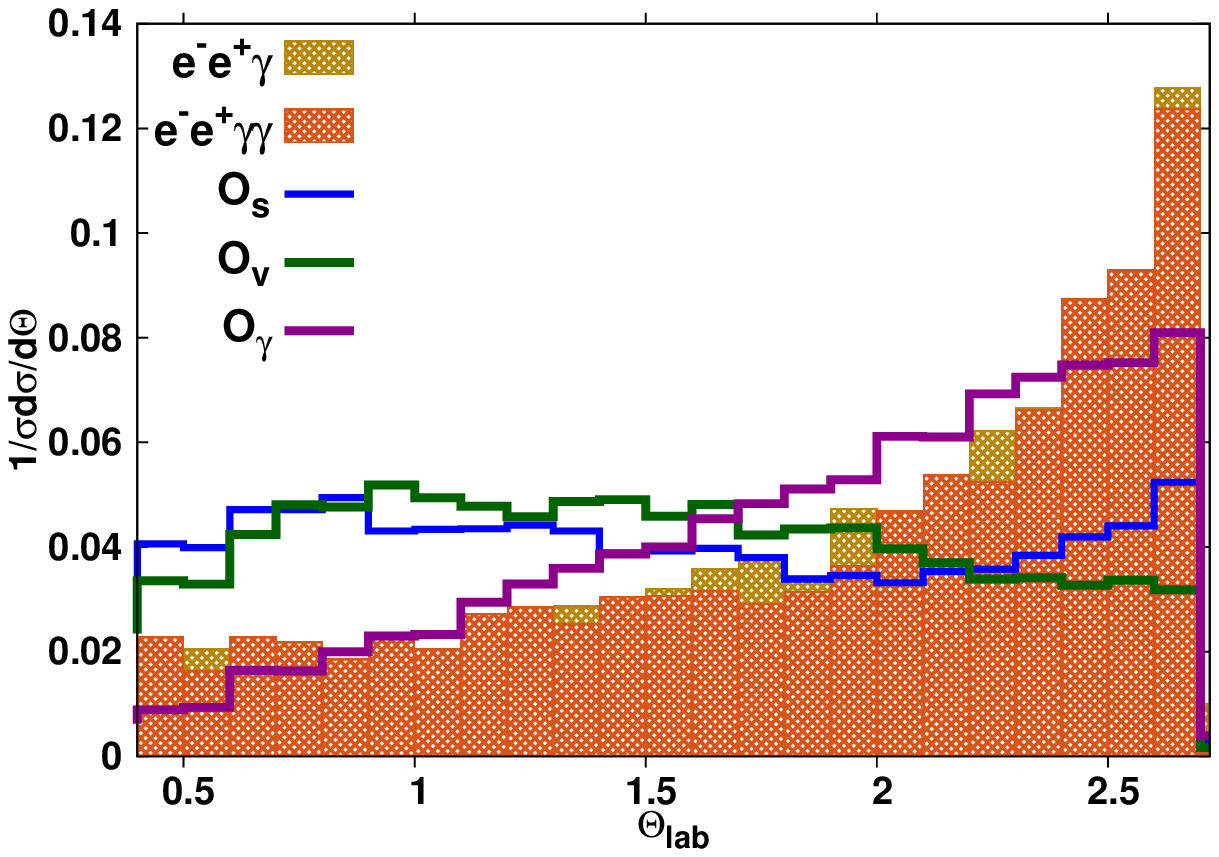}
&  \hskip -05pt
  \includegraphics[width=0.33\textwidth,height=0.33\textwidth,clip]{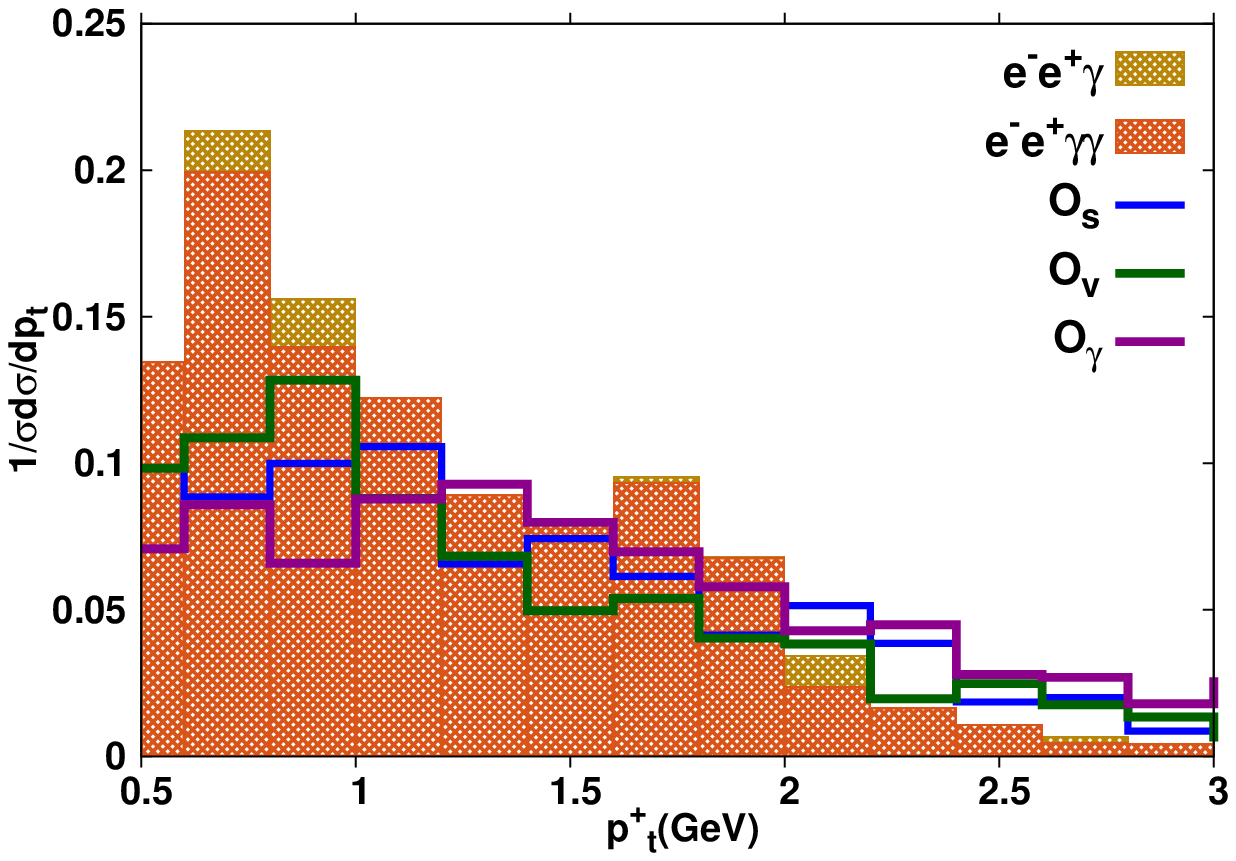}\\
  \hskip -05pt
  \includegraphics[width=0.33\textwidth,height=0.33\textwidth,clip]{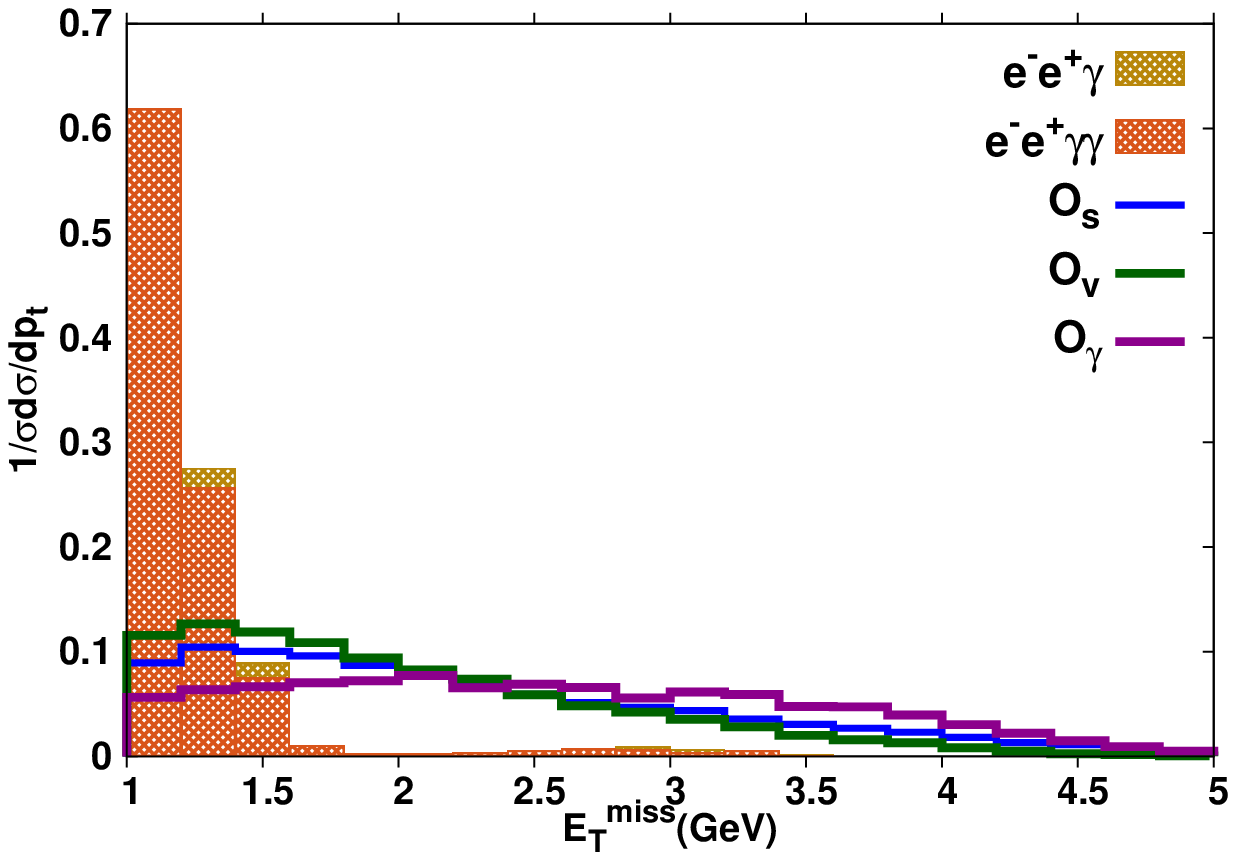} &
  \hskip -05pt
  \includegraphics[width=0.33\textwidth,height=0.33\textwidth,clip]{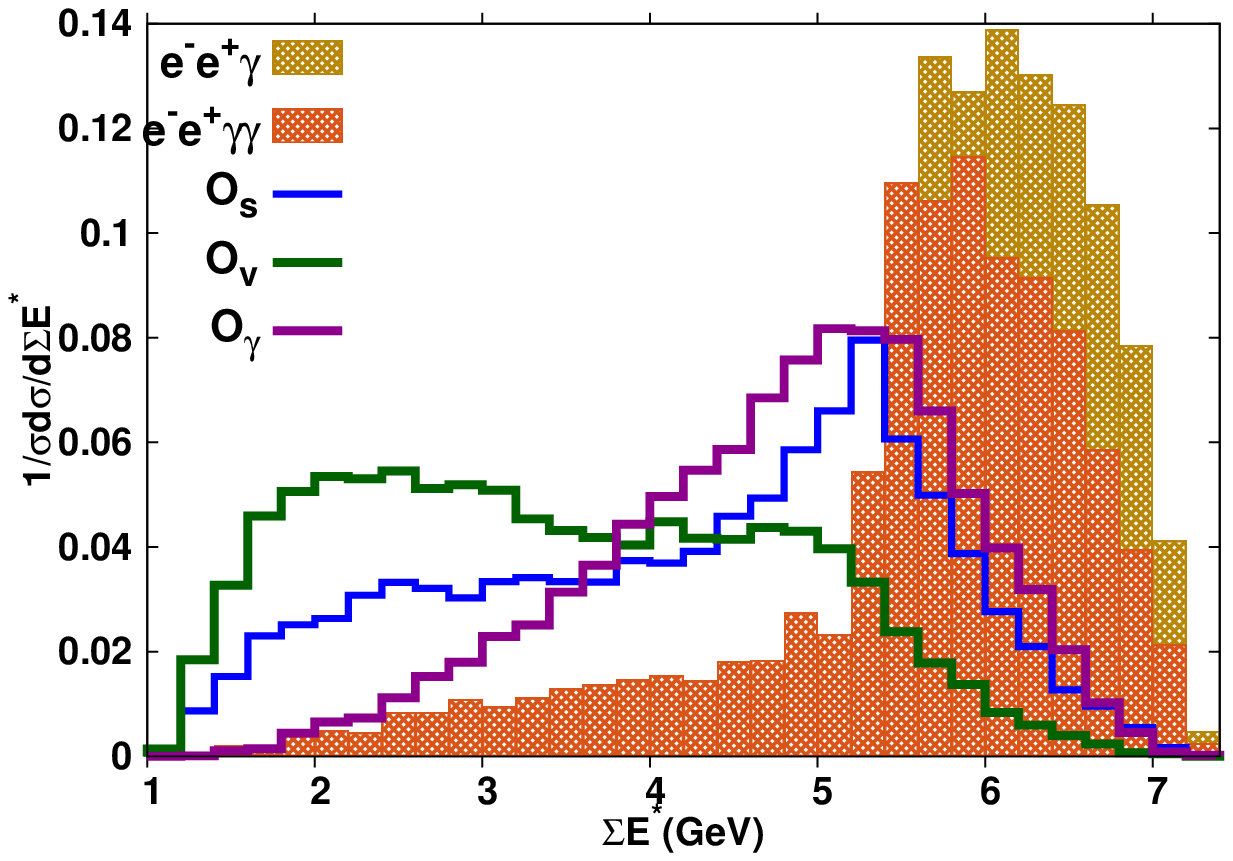} &
  \hskip -05pt
  \includegraphics[width=0.33\textwidth,height=0.33\textwidth,clip]{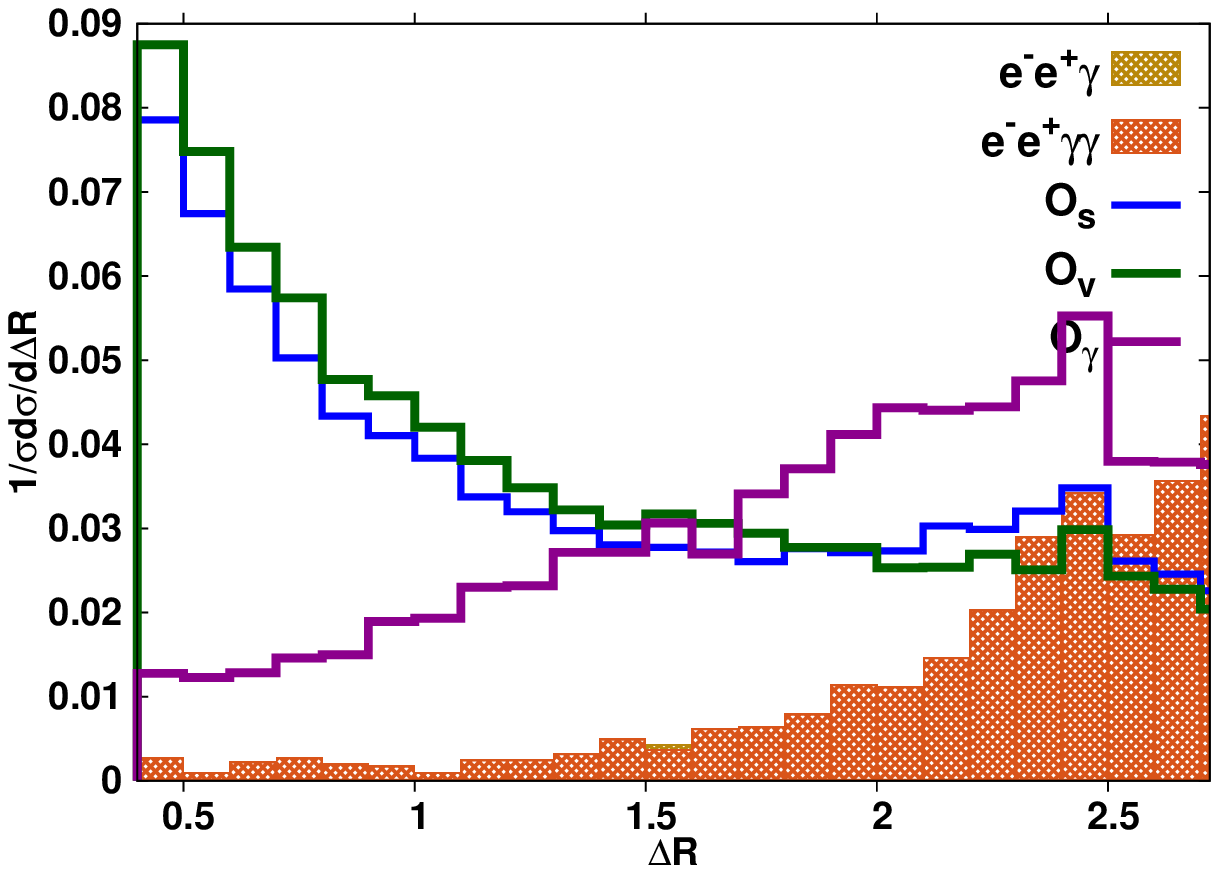}
  \end{tabular} 
  \caption{\small \em Normalized 1-D differential distributions for
    the two major backgrounds as well as signal events, corresponding to $m_\varphi=$ 100 MeV and 
    $\Lambda = 100$ GeV, contributing to
    the $e^- e^+ + \slashed{E}_T$ final state on imposition of the
    basic cuts alone. {\em (a--c)} c.o.m energy, scattering angle and
    transverse momentum of the electron; {\em (d--f)} same for the
    positron; {\em (g)} the missing transverse momentum, {\em (h)} sum
    of the lepton energies and {\em (i)} angular distance $\Delta
    R_{ee}$.}
\label{dist:ee} 
\end{figure}
\item As for the signal events induced by the fermionic
  operators, these can be thought of as $e^- e^+ \to e^- e^+ X$ where
  $X$ denotes a pseudoparticle of variable mass
  $m_{\varphi\varphi}$. Consequently, the kinematics would, to a large
  extent, be analogous to that for the background. However the larger
  effective mass of the $X$ ameliorates the strong forward-backward
  peaking to a significant degree. As for the differences between
  ${\cal O}_{s,p}$ on the one hand and ${\cal O}_{v,a}$ on the other,
  these can be traced to the tensorial nature of the operator
  corresponding to the pseudoparticle.

\item Of particular interest is the distribution of the cone angle
  $\Delta R$ between the $e^\mp$ momenta, defined, in terms of their
  separation in pseudorapidity and in azimuthal plane, as $(\Delta
  R)^2 = (\Delta \eta)^2 + (\Delta \phi_{\rm azim})^2$.  As
  Fig. \ref{dist:ee} shows, the background is concentrated at larger
  values of $\Delta R$, owing directly to their radiative origin. The
  signal events, though, are dominated by events with large
  $m_{\varphi\varphi}$ which, in turn, forces the $e^\mp$ to be
  relatively closer.  Therefore we can profitably use this feature to
  enhance the single to background ratio.

The same kinematic feature is also played up, to a smaller degree, in
the distributions for the missing transverse momentum or the sum of the
lepton energies.

\end{itemize}

\subsubsection{Case II : the $\mu^- \mu^+ + \slashed{E}_T$ final state}

\begin{figure}[H]
  \centering
  \begin{tabular}{ccc}
  \hskip -05pt
  \includegraphics[width=0.33\textwidth,height=0.33\textwidth,clip]{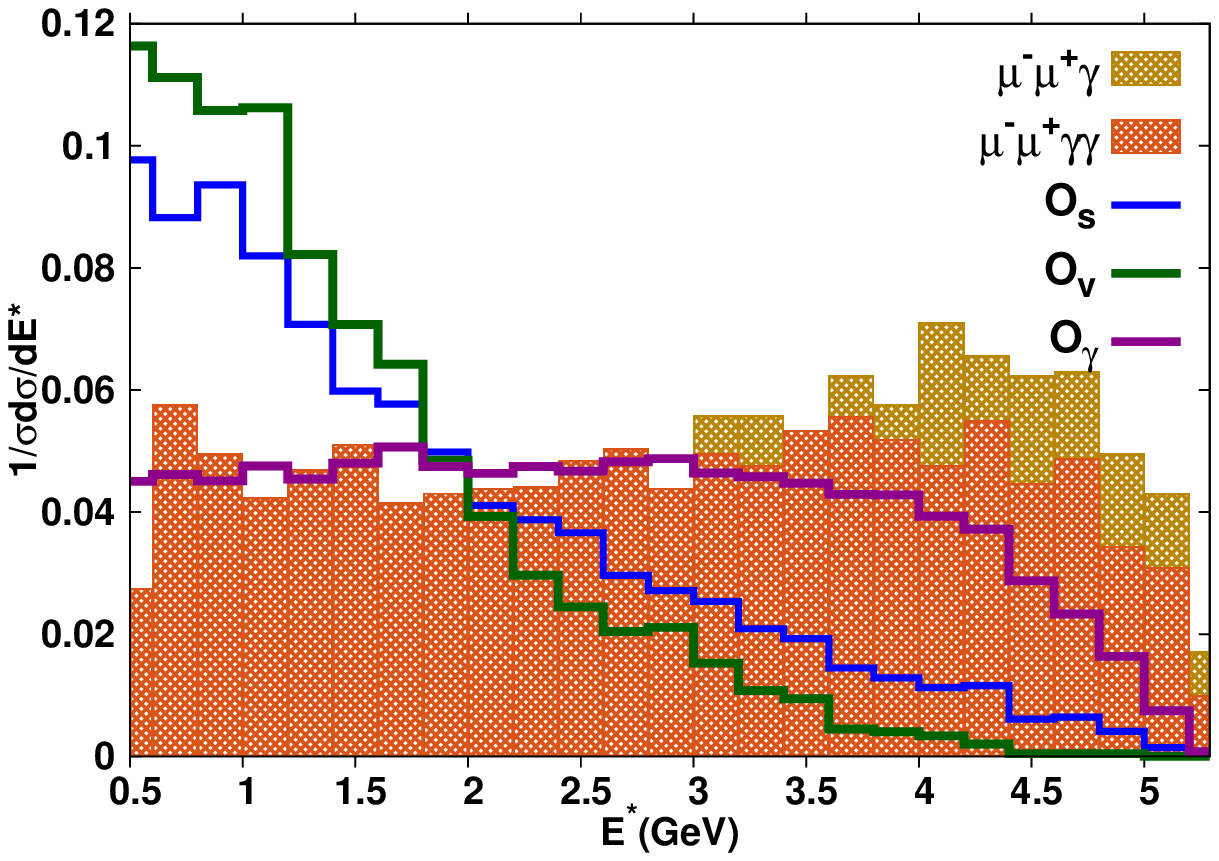} 
&  \hskip -05pt
  \includegraphics[width=0.33\textwidth,height=0.33\textwidth,clip]{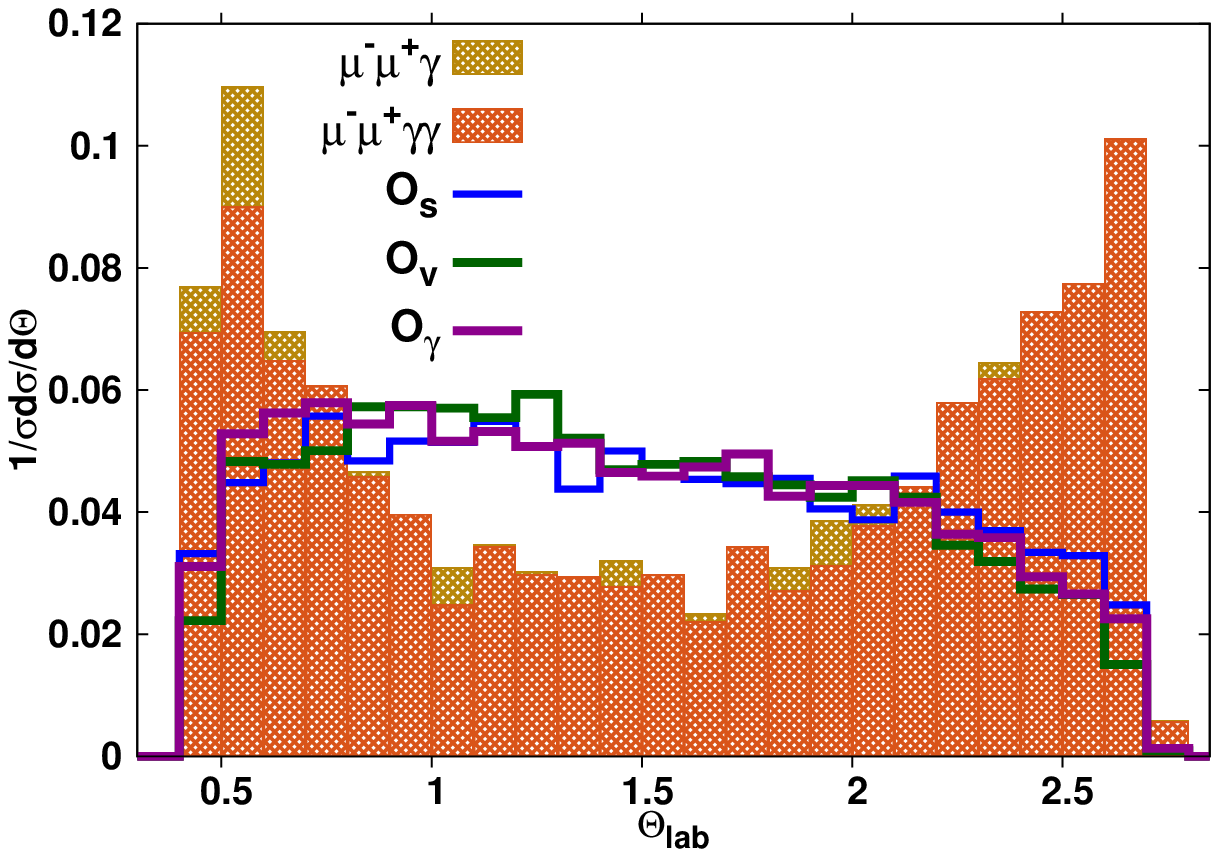} 
&   \hskip -05pt
  \includegraphics[width=0.33\textwidth,height=0.33\textwidth,clip]{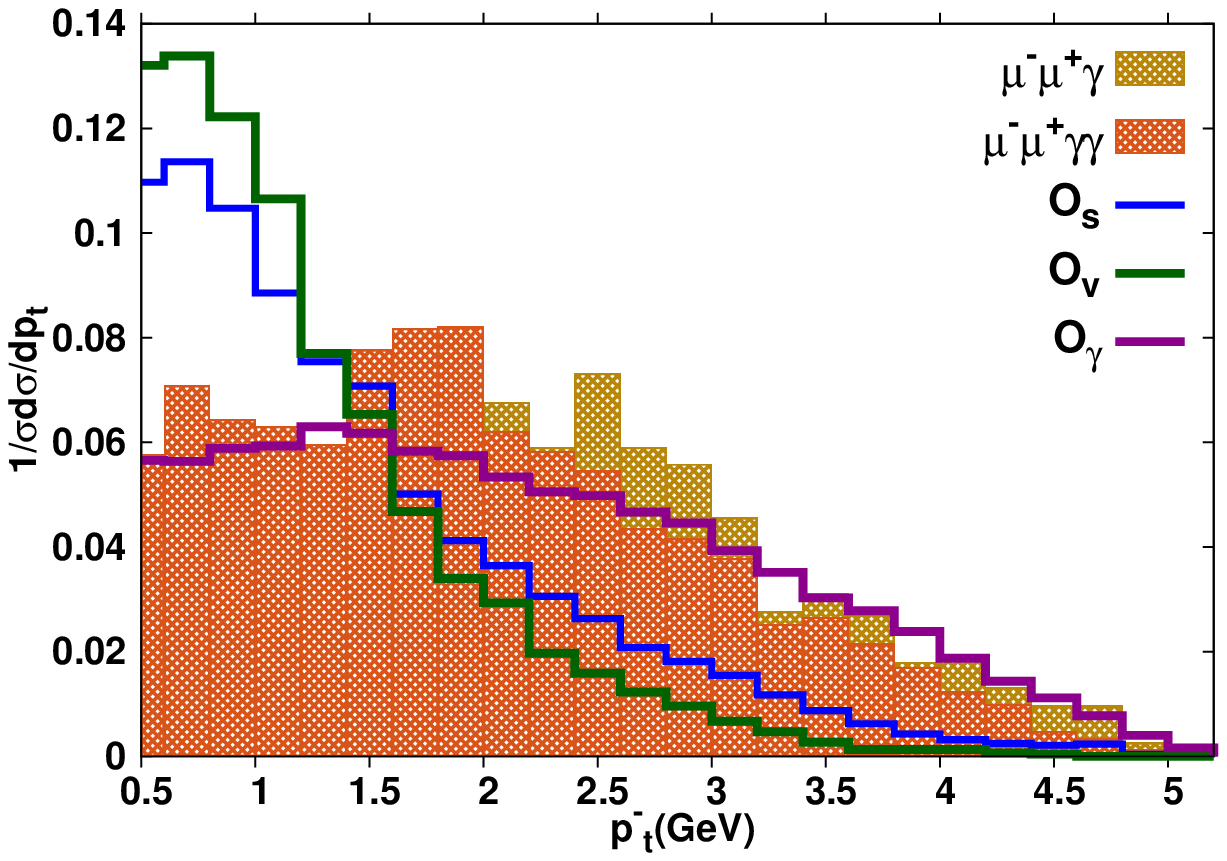} 
\\
  \hskip -05pt
  \includegraphics[width=0.33\textwidth,height=0.33\textwidth,clip]{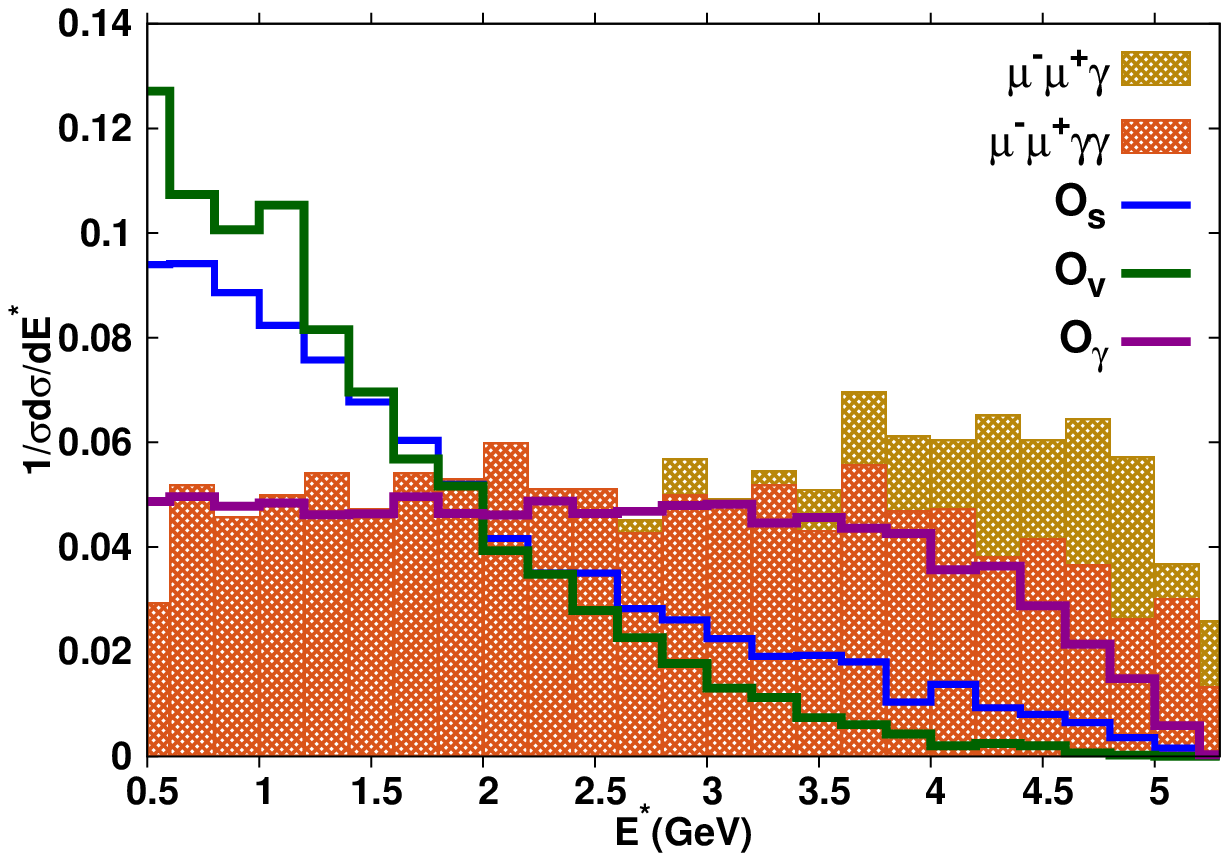} 
&   \hskip -05pt
  \includegraphics[width=0.33\textwidth,height=0.33\textwidth,clip]{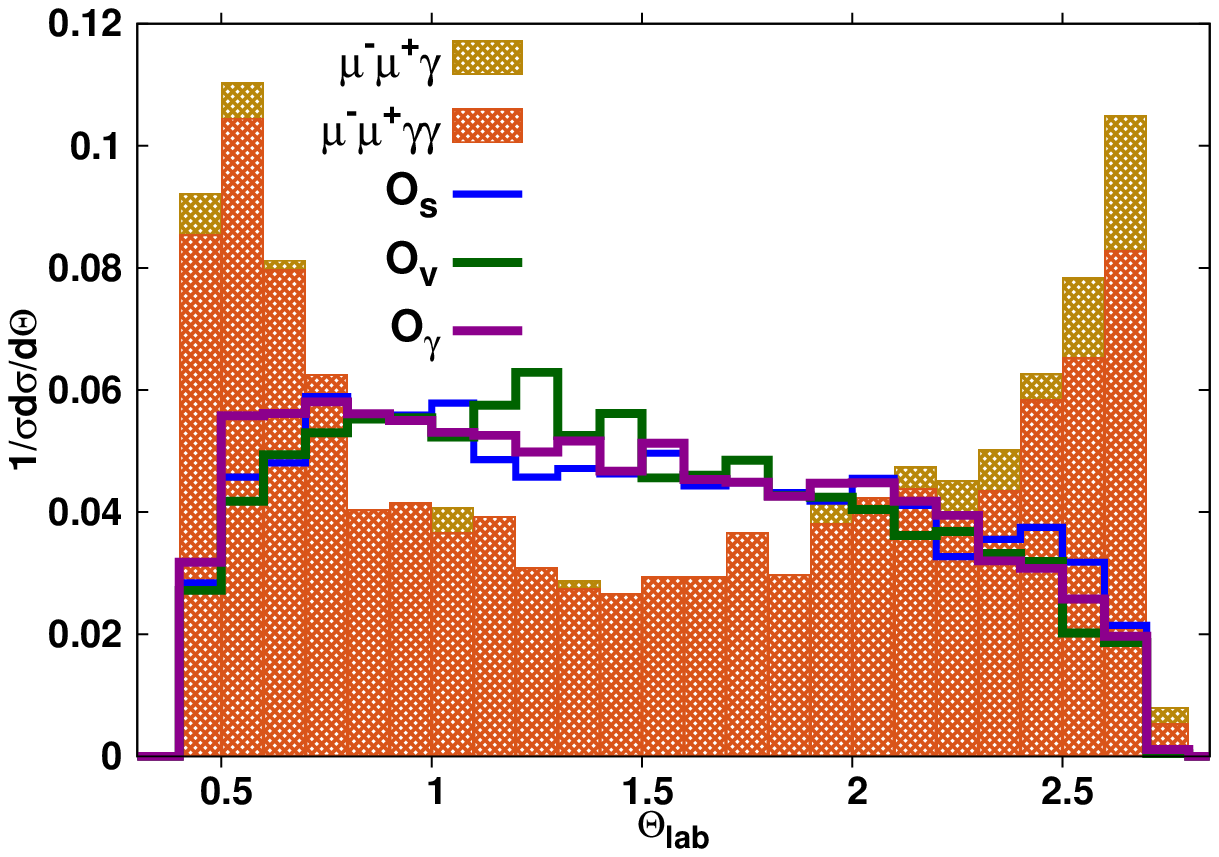}
&  \hskip -05pt
  \includegraphics[width=0.33\textwidth,height=0.33\textwidth,clip]{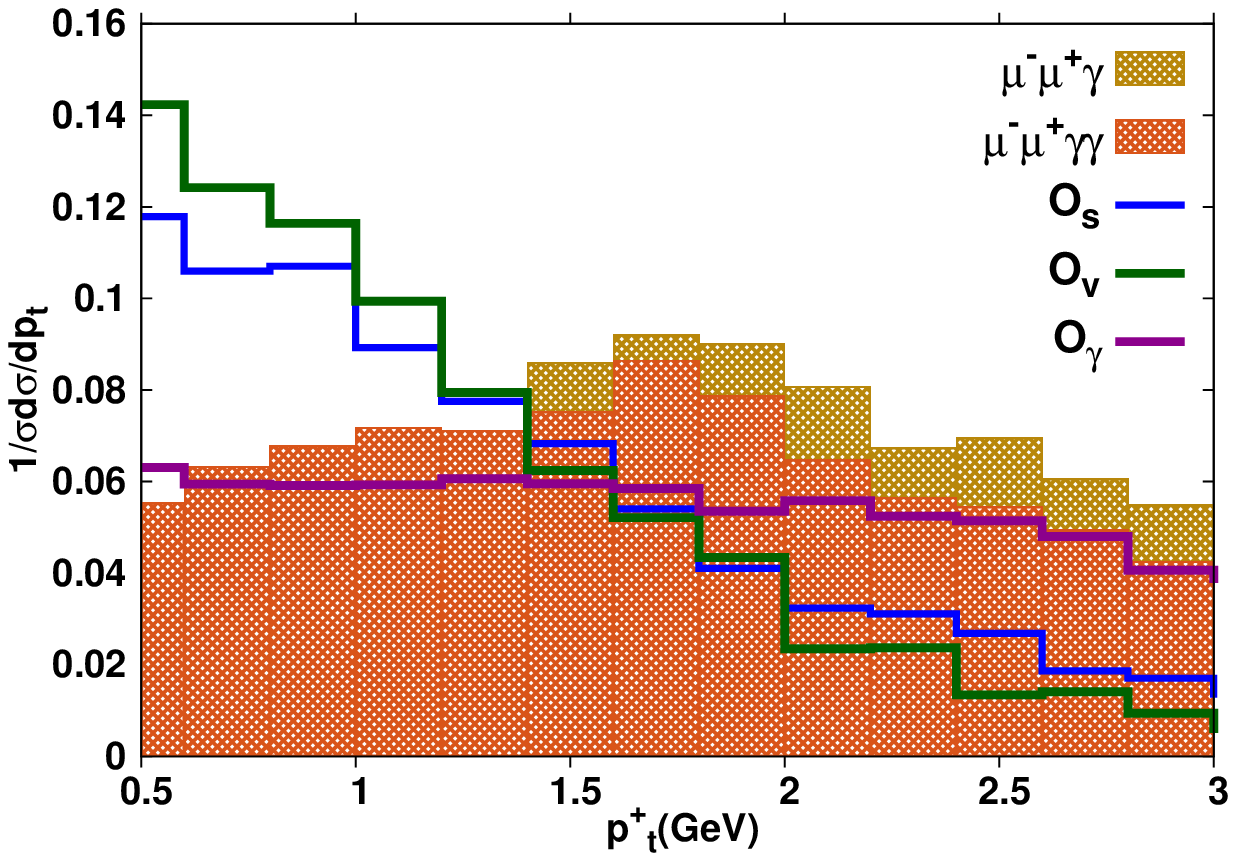}
\\
  \hskip -05pt
  \includegraphics[width=0.33\textwidth,height=0.33\textwidth,clip]{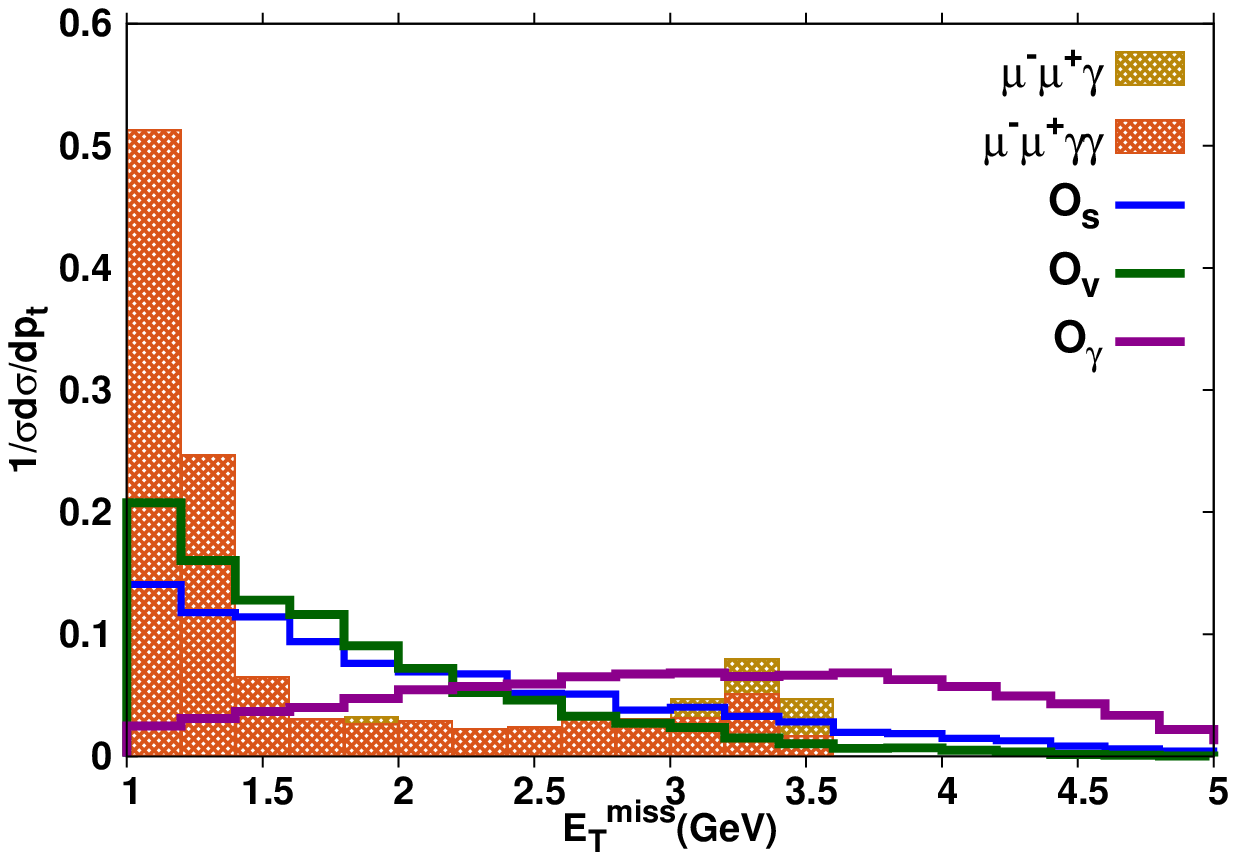} 
&  \hskip -05pt
  \includegraphics[width=0.33\textwidth,height=0.33\textwidth,clip]{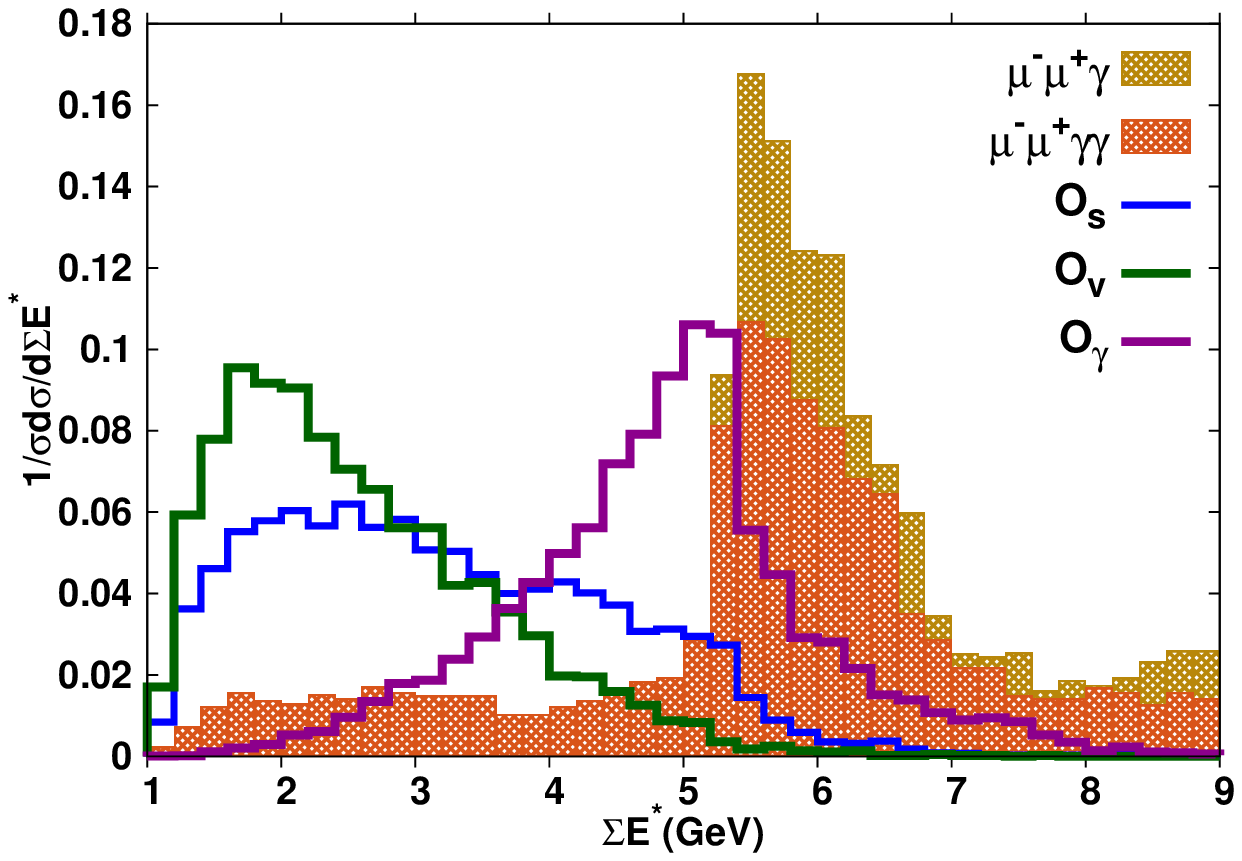} 
&  \hskip -05pt
     \includegraphics[width=0.33\textwidth,height=0.33\textwidth,clip]{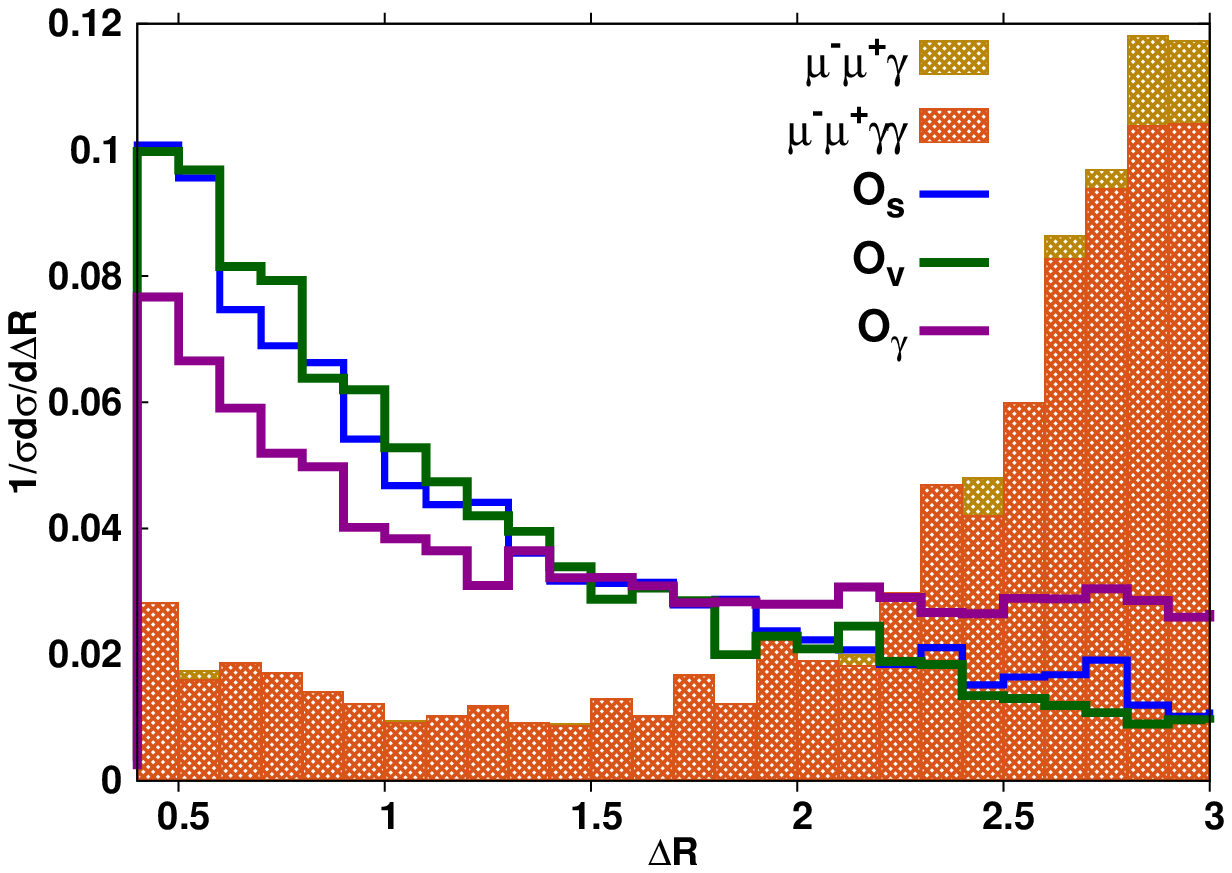}
  \end{tabular} 
  \caption{\small \em As in Fig.\ref{dist:ee}, but for the 
   $\mu^- \mu^+ +  \slashed{E}_T$ final state instead.}
\label{dist:mm}
\end{figure}

\begin{itemize}
\item Despite the lack of the ``$t$-channel-like'' diagrams,
  the distributions for the signal events (Fig.\ref{dist:mm})
  are not very dissimilar from those for the preceding case. This can traced
  to the fact that the kinematic scale, in either case, is being set largely
  by the mass ($m_{\varphi\varphi}$) of the pseudoparticle $X$.

\item For the background events, though, the strong
  forward-backward peaking is ameliorated, leaving behind a muted
  dependence reminiscent of $e^- e^+ \to \mu^- \mu^+$. The remaining
  forward-backward asymmetry is but a consequence of the boosting of
  the center-of-mass frame.

\item Another obvious consequence is the near-identical nature
  of the $E^*$ distributions for the $\mu^\mp$.

\item The preceding arguments (and a look at Fig.\ref{dist:mm})
  suggest that we are faced with a reduced difference between the shapes
  of the signal and background differences, and, hence, reduced sensitivity.
  However, the last row of Fig.\ref{dist:mm} amply demonstrates the fact
  that the differences in the the missing $p_T$, the $\sum E$ and, more particularly,
  the $\Delta R$ distributions persist.
  
\end{itemize}

\subsubsection{Case III : the mono-muon final state}

\begin{figure}[!htb]
  \centering
  \begin{tabular}{ccc}
  \hskip -05pt
     \includegraphics[width=0.33\textwidth,height=0.33\textwidth,clip]{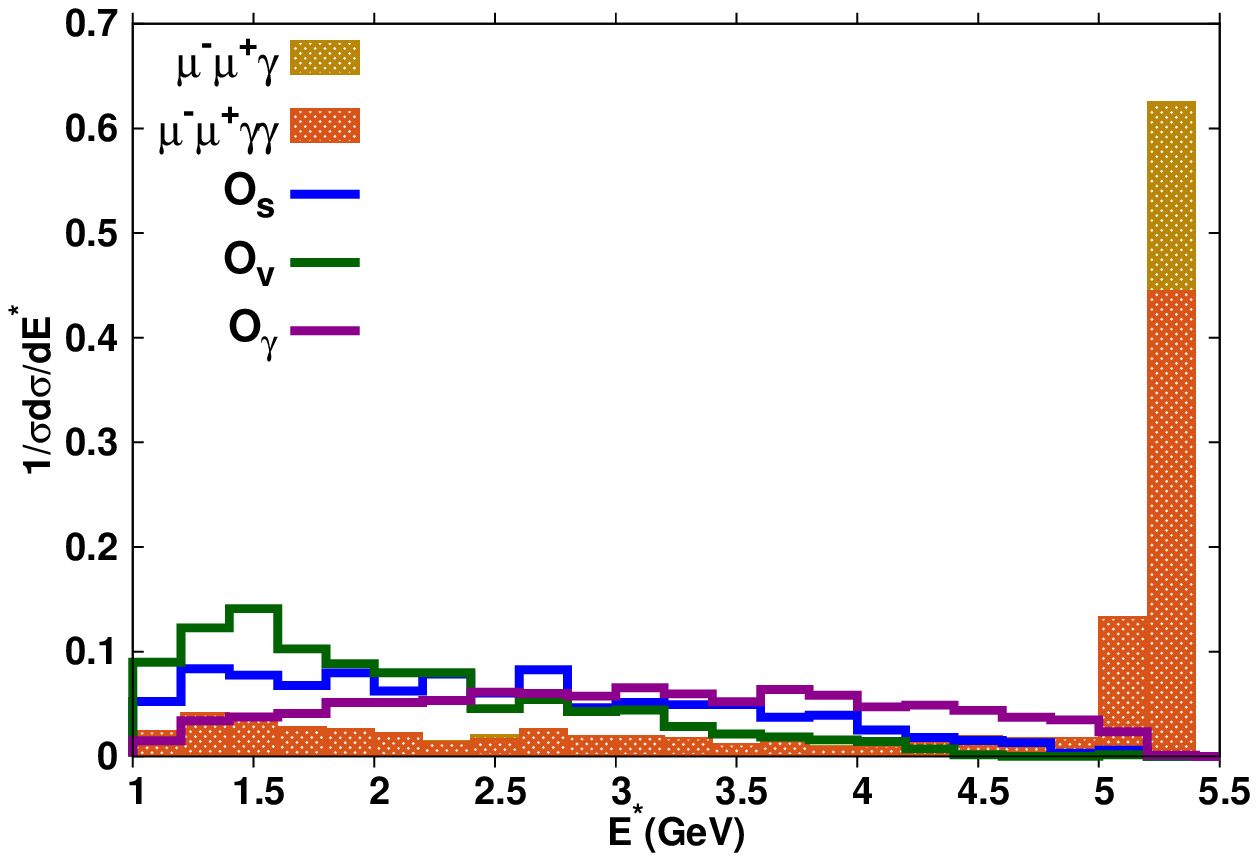} &
  \hskip -05pt
     \includegraphics[width=0.33\textwidth,height=0.33\textwidth,clip]{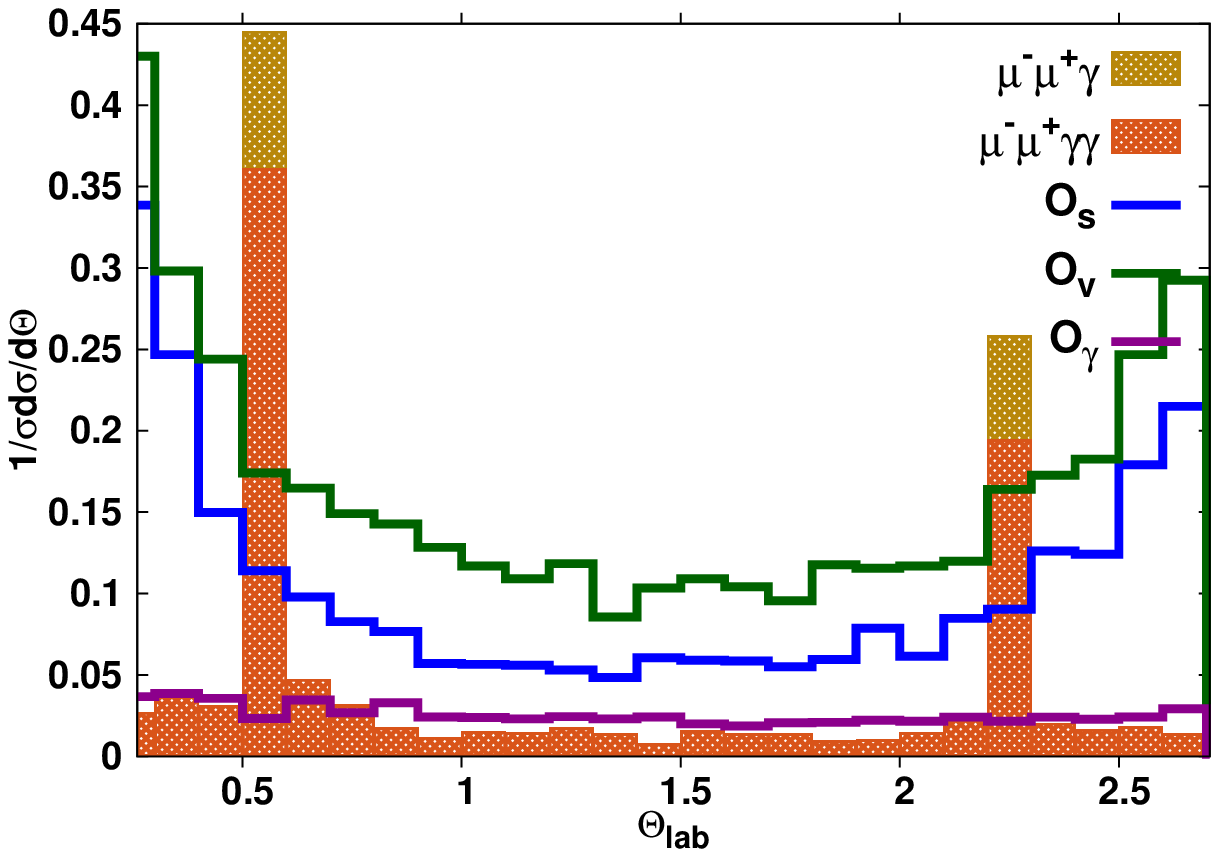}
&
  \hskip -05pt
     \includegraphics[width=0.33\textwidth,height=0.33\textwidth,clip]{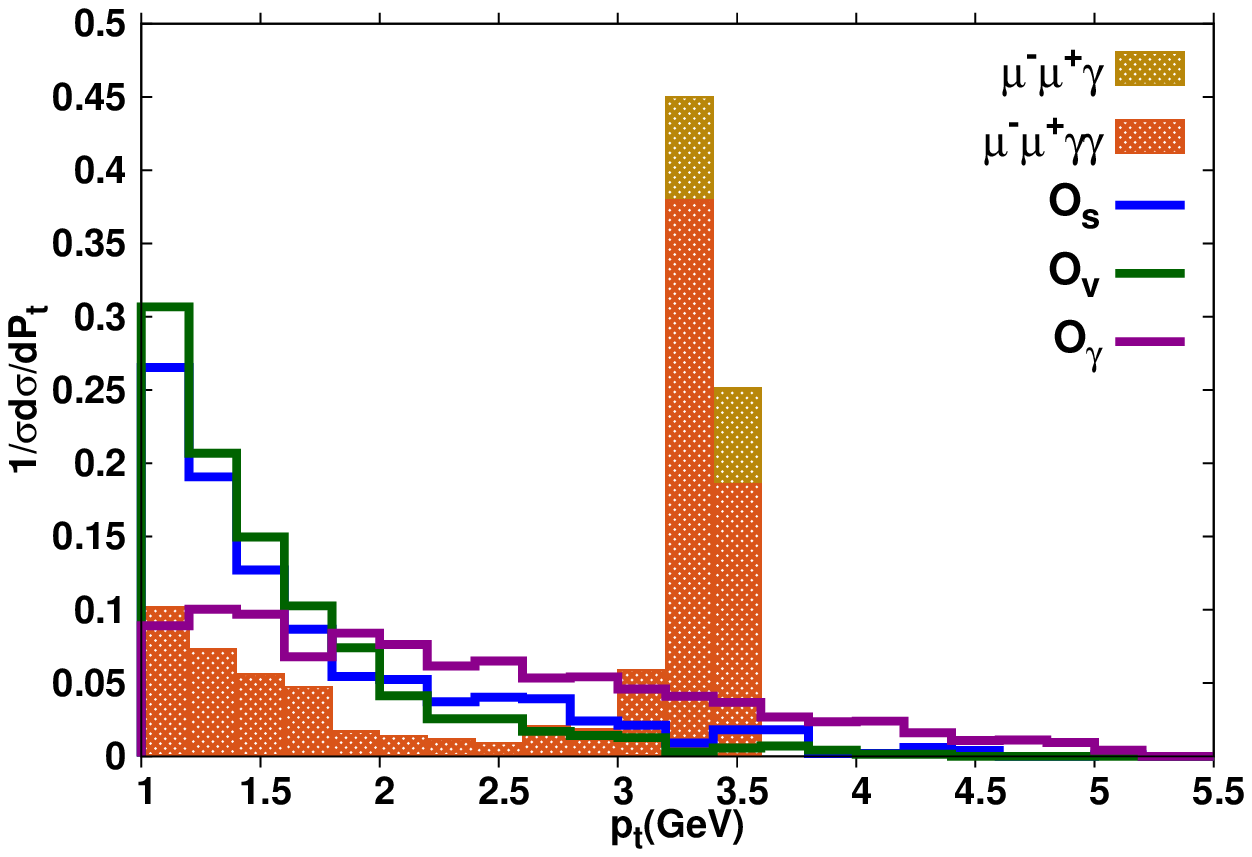}
  \end{tabular}
  \caption{\small \em Normalized 1-D differential distributions for
    the two major backgrounds as well as signal events, corresponding to $m_\varphi=$ 100 MeV and 
    $\Lambda = 100$ GeV, contributing to
    the $\mu^\pm + \slashed{E}_T$ (with only one muon being visible) 
    final state on imposition of the
    basic cuts alone. {\em (a)} c.o.m. energy of the muon 
    {\em (b)} scattering angle in the laboratory frame 
    {\em (c)} transverse momentum.}
\label{dist:mmS} 
\end{figure}

With the basic process being similar to the preceding case, but with
the restriction that only one of the two muons be visible, the
kinematic observables reduce immediately to the same number as in the
mono-photon case. With the kinematics being similar too, there exists
a consequent similarity of the distributions (see Fig.\ref{dist:mmS})
with those for the mono-photon case (as presented in
Sec.\ref{sec:monophot_belle}). That they are are not exactly 
identical is easily understood on realizing that the latter was
dominated by ``$t$-channel-like'' diagrams whereas the present case
has only ``$s$-channel-like'' ones.

\subsection{Selection Cuts and Analysis}
Cases I and II offer us multiple independent kinematic observables
and, hence, the possibility of effecting a detailed multivariate
analysis so as to enhance the signal significance. However, given the
relatively small signal strength, and the level of sophistication of
our event simulation (especially, the treatment of subtle detector
effects), we deliberately desist from adopting such a course. Instead,
we choose to enhance the signal to noise ratio through the imposition
of cuts and, thereafter, attempt a far more conservative analysis of
the data.

 A careful perusal of the distributions motivates us to define the
 selection cuts that we choose: 
\begin{itemize}
 \item Cut-1: An event should contain only one pair of opposite charged 
   leptons with a missing transverse momentum $\slashed{p}_T$ larger than 2 GeV.

 \item Cut-2: The total visible energy should be less than 5 GeV.
 \end{itemize}
\begin{table}[H]
\begin{center}
\begin{tabular}{|c|c|c|c|}
\hline
& \multicolumn{3}{|c|}{Cross sections in pb}\\
\cline{2-4}		
Underlying process & Basic Cuts & $p_T$ > 2  & $\sum E^*$ < 5 \\
\hline
$e^-\, e^+\, \slashed{\gamma}$  & 55	& 12	& 0 \\
$e^-\, e^+\, 2\slashed{\gamma}$ & 17.4& 4.8&$10^{-2}$\\
$e^-\, e^+\, \varphi^\dagger \varphi$ (S) &8$\times 10^{-3}$&4.4$\times 10^{-3}$&2.5$\times 10^{-3}$ \\
$e^-\, e^+\, \varphi^\dagger \varphi$ (V) &1.9$\times 10^{-5}$&8.9$\times 10^{-6}$&6.6$\times 10^{-6}$\\
\hline
\hline
$\mu^- \, \mu^+\, \slashed{\gamma}$& 2.3& 0.4&0\\
$\mu^-\, \mu^+\, 2\slashed{\gamma}$& 1.6& 0.1&3.4$\times 10^{-3}$\\
$\mu^-\, \mu^+\, \varphi^\dagger \varphi$ (S)& $10^{-2}$&3.5$\times 10^{-3}$&2.7$\times 10^{-3}$\\
$\mu^-\, \mu^+\, \varphi^\dagger \varphi$ (V)& 3.2$\times 10^{-5}$&6.8$\times 10^{-6}$& 6.3$\times 10^{-6}$\\
\hline
\end{tabular}
\end{center}
\caption{\small \em The dependence of the signal and the 
  two leading background cross
  sections for Case I (upper set) and Case II (lower set) on the
  cuts. The signal has been computed for a representative value of
  $m_\varphi$=100 MeV and $\Lambda=$ 100 GeV}
\label{tab:caseI&II_cutflow}
\end{table}

\begin{table}[H]
\begin{center}
\begin{tabular}{|c|c|c|c|}
\hline
 Single&\multicolumn{3}{|c|}{Cross sections in pb}\\
\cline{2-4}
lepton channel & Basic Cuts & $p_T$ < 2 \& $E^*$ < 5 & $p_T$ > 2 \& $E^*$ < 5 \\
\hline
$\mu \, \slashed{\mu}\, \slashed{\gamma}$& 35& 0&0\\
$\mu\, \slashed{\mu}\, 2\slashed{\gamma}$& 7.5&0.6 & 8.1 $\times 10^{-4}$\\
$\mu\, \slashed{\mu}\, \varphi^\dagger \varphi$ (S)& 8.6 $\times 10^{-3}$ &8.2 $\times 10^{-3}$ &4.3 $\times 10^{-4}$\\
$\mu\, \slashed{\mu}\, \varphi^\dagger \varphi$ (V)& 3.7 $\times 10^{-5}$&3.6$\times 10^{-5}$ &7 $\times 10^{-7}$\\
\hline
\end{tabular}
\end{center}
\caption{\small \em The dependence of the signal and the 
  two leading background cross sections for Case III on the
  cuts. The signal has been computed for a representative value of
  $m_\varphi$=100 MeV and $\Lambda=$ 100 GeV}
\label{tab:caseIII_cutflow}
\end{table}
While the background process with a single hard photon, naturally,
has, to start with, a much larger cross section than that with two
hard photons, it suffers more severely due to the cuts (see Table
\ref{tab:caseI&II_cutflow} \& \ref{tab:caseIII_cutflow}).  This is easy to understand as the
balancing of the leptonic $p_T$ and, even more, the missing energy by
a single photon, makes it very difficult for the latter to have
escaped the detector by going either sufficiently forward or
backward. Thus, it has to, essentially, fall into the ECL cracks. On
the other hand, if the missing momentum were to be
shared by two photons, there is a much higher probability of the event
satisfying the selection cuts. It might seem, at this stage, that even
higher order processes such as $e^- e^+ \to \ell^- \ell^+ + 3
\slashed{\gamma}$ could contribute non trivially to the
background. This, however, is not so. The addition of one further
photon does not make it any easier for the photons to have all missed
detection as well as satisfy the two selection cuts. Rather, with the
additional suppression by a factor ${\cal O}(\alpha_{\rm em})$, the
ensuing contribution is actually much smaller.

\begin{figure}[!htb]
\begin{center}
\includegraphics[width=0.4\textwidth,clip]{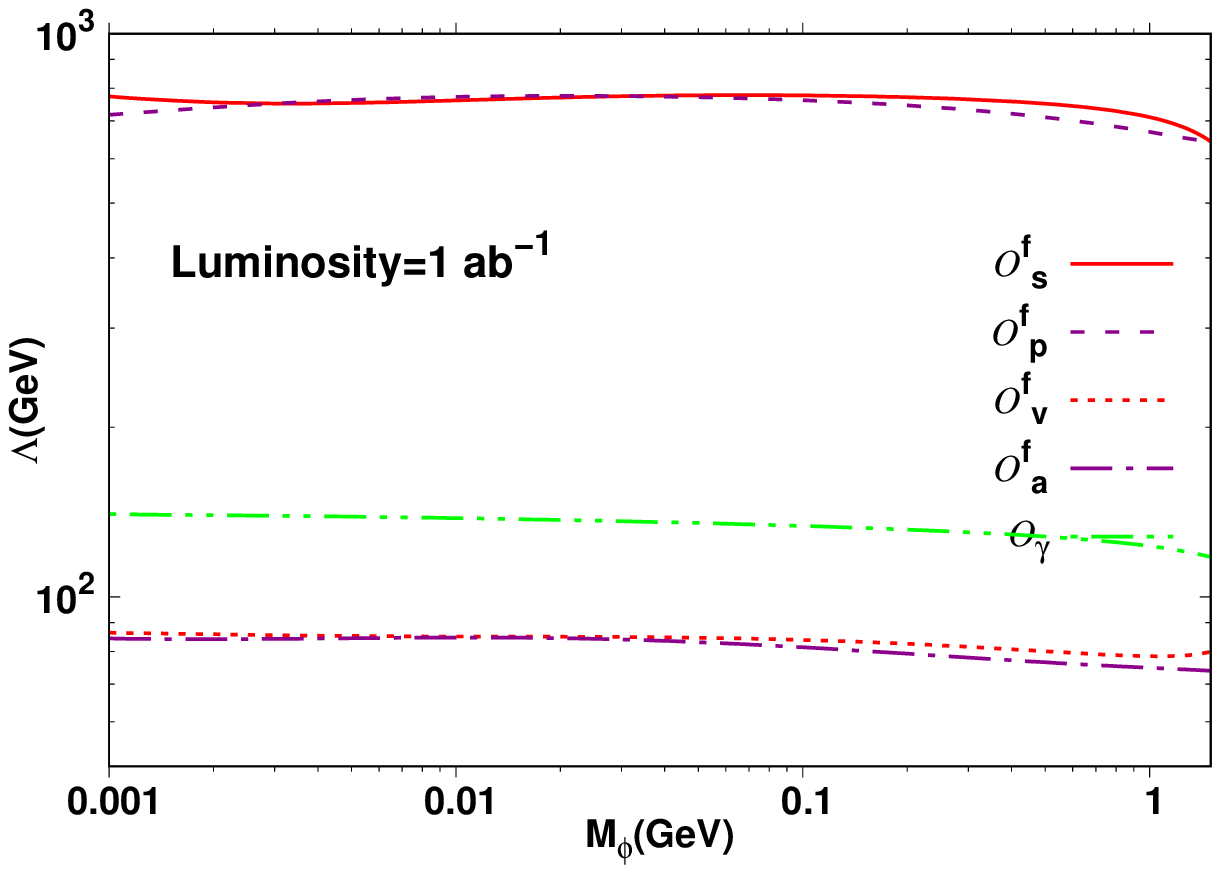}
\includegraphics[width=0.4\textwidth,clip]{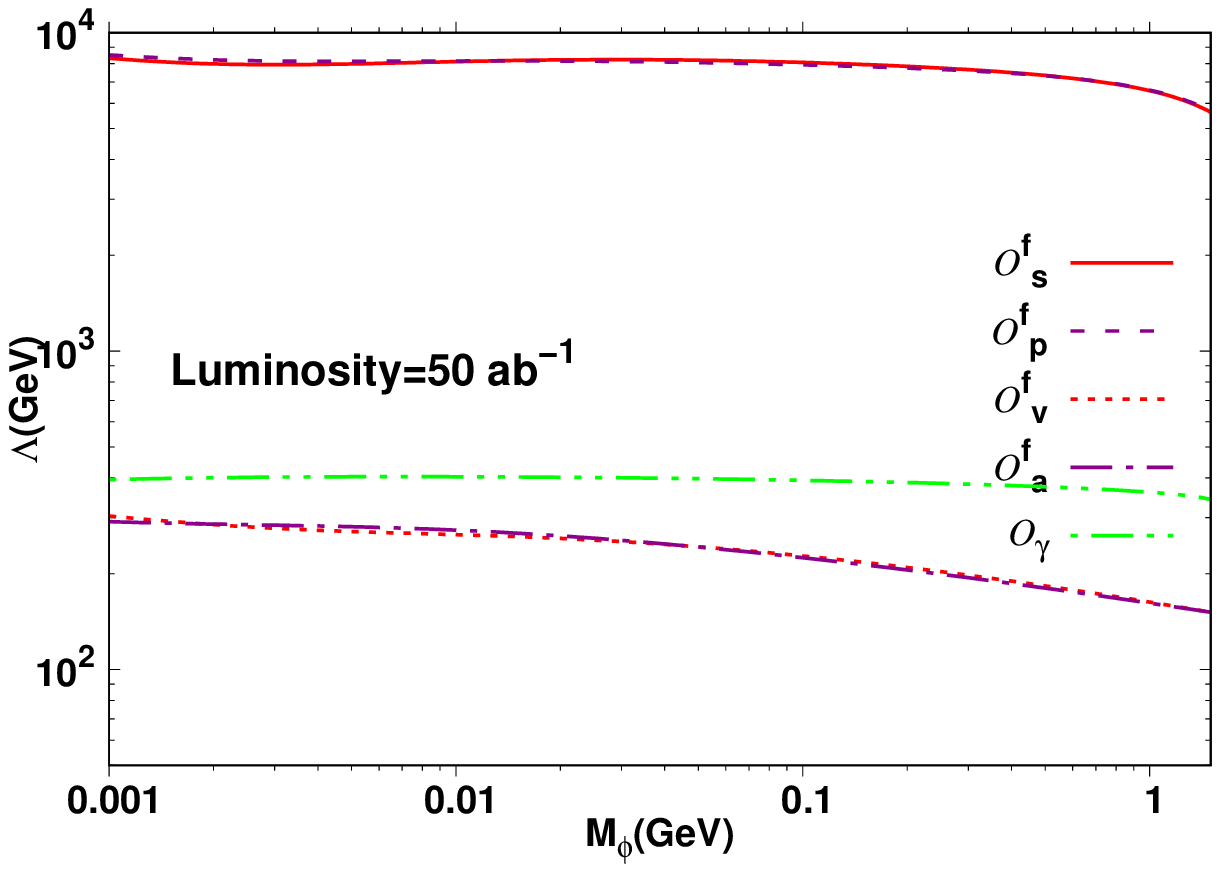}

\includegraphics[width=0.4\textwidth,clip]{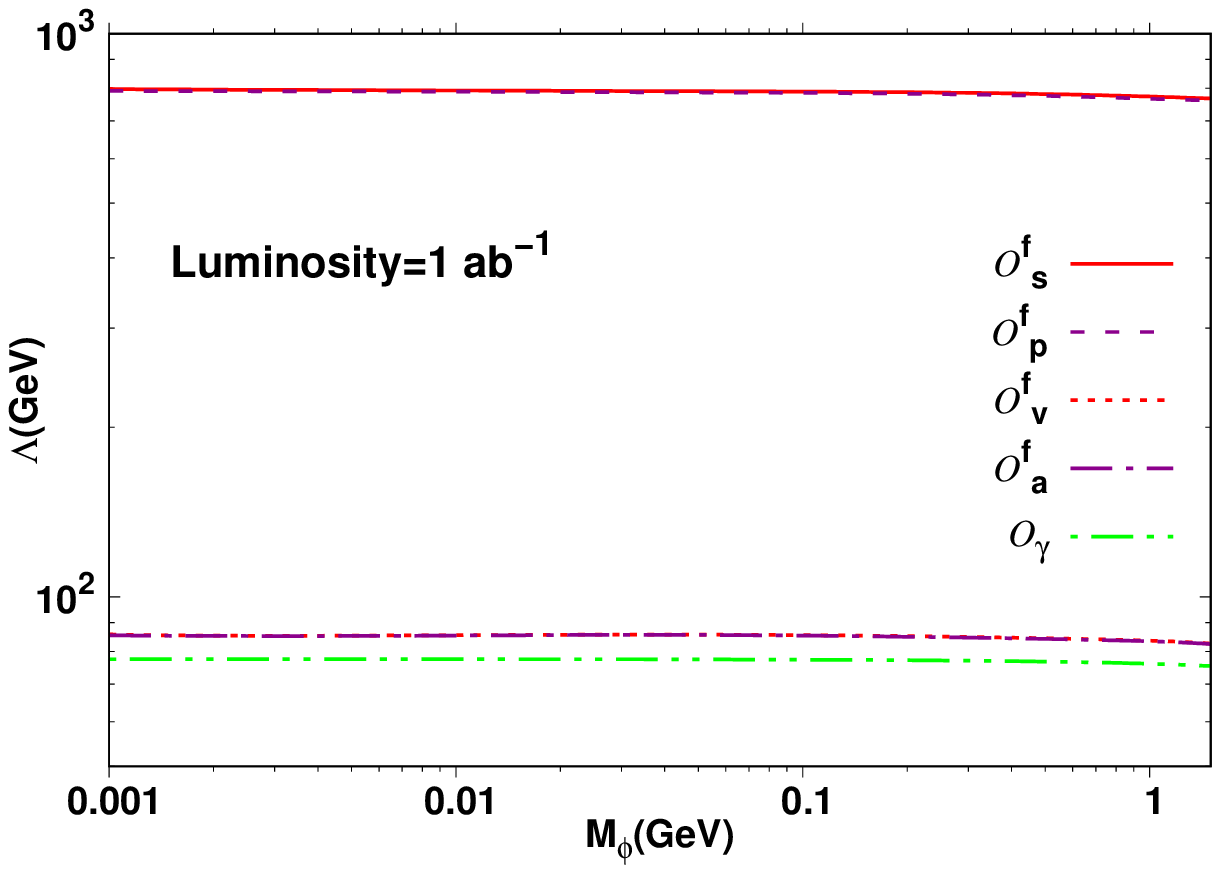}
\includegraphics[width=0.4\textwidth,clip]{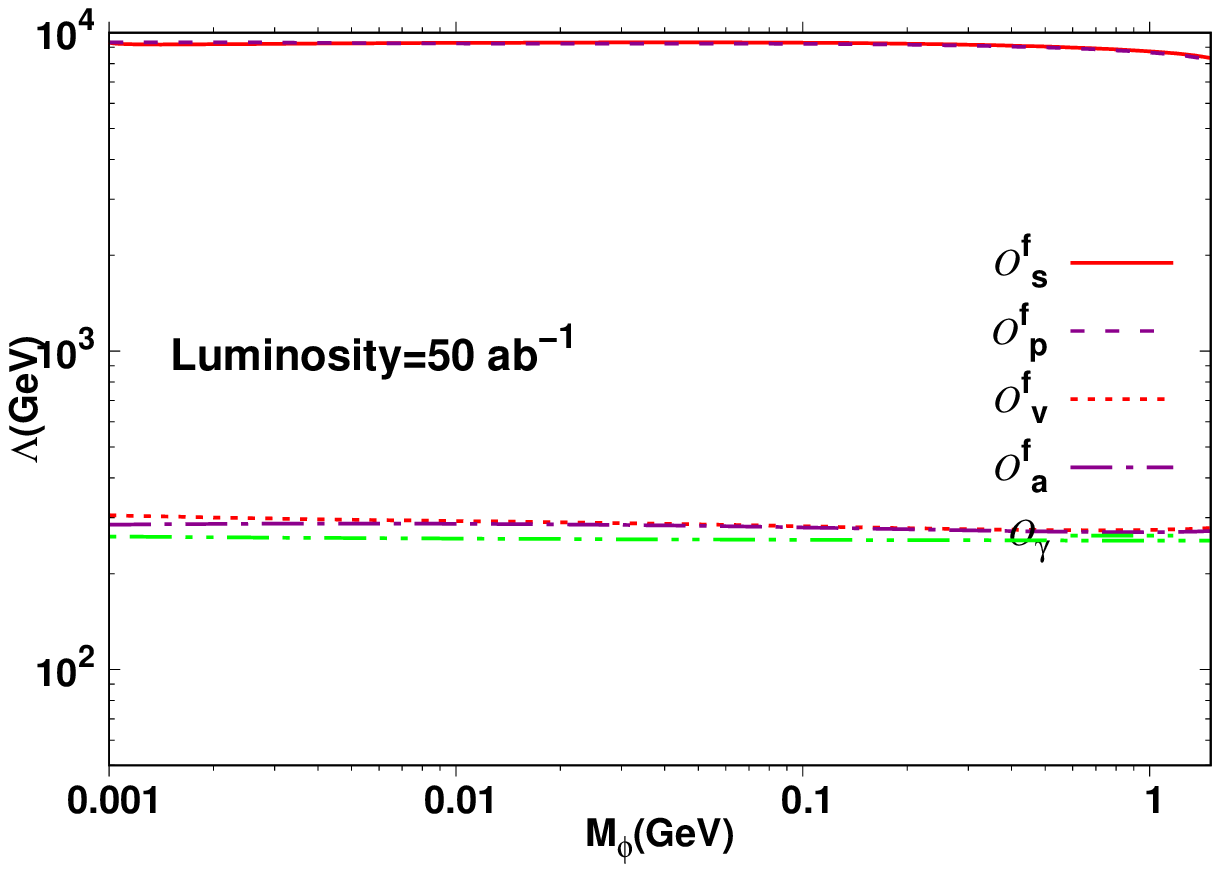}
\end{center}
     \caption{\small \em 99\% C.L. exclusion contours in the $m_{\varphi}$--$\Lambda$ obtainable, at Belle-II, from analyses of final states comprising a dilepton with 
missing energy. Upper (lower) sets correspond to $e^-e^+ \, (\mu^-\mu^+)$
respectively, while left and right panels assume an integrated luminosity of 
 $1 \& 50 \,{\rm ab}^{-1}$ respectively.}
\label{fig:contour_I&II}
\end{figure}

Having imposed the aforementioned selection cuts, we now constitute a
uniform two dimensional grid in the $\Delta R$--$\sum E_\ell$
plane. Comparing the signal and background event count in each of the
 bins (sized $0.1 \times 0.2 \gev$), we perfrom a $\chi^2$ test. The consequent
exclusion contours are presented in Fig.\ref{fig:contour_I&II}. Note
the large improvement in sensitivity going from an integrated
luminosity of $1 \, {\rm ab}^{-1}$ to $50 \, {\rm ab}^{-1}$, an
improvement much larger than the corresponding
obtainable in the case of the monophoton signal. This only reflects
the fact that, for the smaller luminosity, these processes are
statistics-limited, whereas the monophoton signal quickly became
systematics-limited.

While the sensitivities, as shown by Fig.\ref{fig:contour_I&II} are
systematically lower than that available from the monophoton case,
note that the suppression factor is not too large, especially for the
large luminosity case. Thus, these channels do serve to provide
additional information. It is obvious that combining the two
  channels would lead to even better constraints. Case III, on the
other hand, shows weaker sensitivity.

\section{Additional Remarks}
   \label{sec:additional}
We now comment on the apparent lack of full $SU(2) \otimes U(1)$ 
symmetry of the ${\cal O}^f_{s,p}$ operators. As has been remarked earlier,
this symmetry can be restored if we consider, instead, operators of the form
\begin{equation}
{\cal O}^f_{s} \to \frac{\xi_f}{\Lambda^2} \; 
    H_{\rm SM} \; \varphi^\dagger\varphi   \; \bar{f} f 
\label{dim_6_scalar}
\end{equation}
and, analogously, for 
${\cal O}^f_{p}$. Here $H_{\rm SM}$ is the SM Higgs field. Post electroweak 
symmetry breaking, the relevant piece of the Lagrangian can be written as
\[
{\cal O}^f_{s} \supset \frac{\widehat{\mathcal{C}^f_{s}}}{\Lambda} \; 
     \varphi^\dagger\varphi   \; \bar{f} f \ , 
  \qquad 
\widehat{\mathcal{C}^f_{s}} \equiv \xi_f \, \frac{v}{\Lambda} \ ,
\]
  where $v = \langle H_{\rm SM}^0 \rangle$ is the symmetry breaking
  scale. If we {\it assume} that $\xi_f$ is comparable to the usual
  fermion Yukawa coupling, $\widehat{\mathcal{C}^f_{s}}$ would be tiny
  for the light fermions, rather than ${\cal O}(1)$ as we have
  assumed.  Consequently, the sensitivity to $\Lambda$ reduces
  enormously.  Indeed, sensitivity to $\Lambda > 10 \gev$ (a must for
  the effective theory paradigm to be valid at Belle) requires an
  integrated luminosity $\gapp 8 \, {\rm ab}^{-1}$. While
  ref.\cite{Yu:2016pqp} claims a much better sensitivity for low
  luminosities, note that they seek to benefit from an enhanced
  coupling to the charm quark by looking at $e^+ e^- \to \varphi +
  \varphi^* + J/\psi$ or $e^+ e^- \to \varphi + \varphi^* +
  \eta_c$. Nonetheless, for an integrated luminosity of $50 \, {\rm
    ab}^{-1}$, the sensitivity of our (monophoton) mode is only a
  factor of $\sim 1.5$ worse than Ref.\cite{Yu:2016pqp},  
  and hence this mode constitutes
  an important additional probe. Note, further, that the $J/\psi$ (or
  $\eta_c$) modes are kinematically inaccessible for $m_\varphi \gapp
  3 \gev$, and the monophoton mode would be the best bet for such
  masses.

  Since the reduced sensitivity is a consequence of our having
  assumed that $\xi_f$ are of the order of the usual Yukawa couplings,
  it is interesting to consider the opposite case of $\xi_f \sim {\cal
    O}(1)$ instead. Such a situation could transpire if the operator
  of eq.(\ref{dim_6_scalar}) were the result of some strong
  dynamics. For such a case, the consequent bounds on $\Lambda$ would
  be only marginally weaker than those on the corresponding
  dimension-5 operator of eq.(\ref{the_operators}). A more interesting
  outcome of such a value for $\xi_f$ would be a four-body decay of
  the form
\[
H^0 \to \varphi + \varphi^* + f + \bar f \ .
\] 
  While these rates are far smaller than those for the two-body
  decay modes, they are significantly larger than the SM rates for
  $H^0 \to \nu_i + \bar\nu_i + f + \bar f$. Although they are still
  too small to have been identified at the LHC (with a further
  experimental complication on account of the spread in the invariant
  mass of the $f \bar f$ pair), it would be interesting to look for these
  as the integrated luminosity mounts.

Finally, if $\varphi$ were a real scalar field (rather than a complex
one) with identically defined couplings, the cross sections would be
larger by a factor of two (owing to there being two identical
particles in the final state). Consequently, for a given $m_\varphi$, the
constraints on $\Lambda_{{\rm real} \varphi}$ would be a factor of
$\sqrt{2} \, (2^{1/4})$ stronger for the analogues of operators
$\mathcal{O}^f_{s,p}$ ($\mathcal{O}^f_{\gamma,\tilde{\gamma}}$).

\section{Direct Detection} 
   \label{sec:direct_detect}
The ambient (in the vicinity of the earth) density of DM can, in
principle, be probed through their interaction with terrestrial
detectors. The sensitivity, of course, would be dependent not only on
the experimental configuration, but also on the profile of the DM
distribution in the immediate neighborhood. Several such profiles, 
  defined not only in terms of the density but also in terms of
velocity, have been extensively studied in the
literature~\cite{Navarro:1995iw,2006AJ....132.2685M}.  Fortunately
though, the consequent differences are stark only close to the
galactic center, while at the periphery, the experimental expectations
are quite similar. Consequently, for the rest of the section, we will
make use of the following standard assumptions: a Maxwell-Boltzmann
velocity distribution with its high-velocity tail truncated at the
galactic escape velocity of 544 km/s, a local velocity dispersion of
the DM halo, $v_{rms}=270$ km/s, and a local dark matter density of
0.3--0.4 GeV/cm$^3$.

Such Direct Detection experiments, typically, involve the use of a
bolometric device working at a very low temperature (so as to
eliminate thermal noise to the maximum extent possible). With the
WIMP-nucleus cross sections being much larger than WIMP-electron ones
(for similar-sized couplings), these experiments are much more
sensitive to the former, to the extent of often neglecting bounds on
the latter. In particular, nucleon
collision cross sections for a DM with a mass in the 0.5--5 GeV range
are severely constrained by the
CRESST-II~\cite{Angloher:2015ewa} experiment.  On the other hand, for
masses below 1 GeV, the typical recoil energy is lower than the
detector thresholds, rendering such experiments quite insensitive.

In the case DM is hadrophobic, its interaction with the detector
material would proceed primarily through its interaction with the
electrons therein and this is what we would start this section
with. However, we shall end our analysis with the study of nucleon-DM
interaction, which will be relevant if DM interacts with quarks also.

\subsection{DM scattering off electrons} 

Even for masses below 1 GeV, DM scattering off electron can lead to a
measurable signal. Consequently, many experiments (including those for
which the primary mode is a different one) have investigated this, for
example, in the context of inelastic electron scattering leading to
ionization of atoms.  For semiconductor targets, excitation of an
electron to above the band gap is also of interest. Sensitivity
to such processes has been studied in context of the XENON10 detector
in Ref.\cite{Essig:2012yx}. The rate for
such processes are dependent on three factors: the ionization form
factor, the elastic WIMP-electron cross section and the density
profile for the DM particle.

To begin with, we consider the leading elastic WIMP-electron 
cross sections for the operators in eq.(\ref{the_operators}). 
These are 
\begin{equation}
\barr{rcl}
\dis  \sigma_{\varphi e}(\mathcal{O}^f_{s})&=&\dis 
  \frac{m_e^2}{4\pi\Lambda^2 (m_{\varphi}+m_e)^2}\\[3ex]
\dis  \sigma_{\varphi e}(\mathcal{O}^f_{v})&=&
\dis \frac{m_e^2 m_\varphi^2}{\pi\Lambda^4(m_\varphi+m_e)^2} \ ,
\earr
    \label{dd_eDM}
\end{equation}
whereas, for the other operators (in eq.\ref{the_operators}) involving
the electron, the cross sections are further suppressed by powers of
the velocity $v$ of the DM. Since, near earth, we only have $v \sim
10^{-3}$ (in units of the velocity of light), the
  corresponding cross sections are negligibly small.

Naively, it might seem that the two operators involving the photon
(namely, $\mathcal{O}_{\gamma}$ and $\mathcal{O}_{\tilde{\gamma}}$)
would trivially escape any constraints emanating from electron-DM
scattering.  However, at the one-loop level, (ostensibly dimension-5)
$\varphi \varphi^* e^+ e^-$ effective operators
$\mathcal{O}^e_{s,p}(\text{eff.})$ corresponding to
$\mathcal{O}_{\gamma,\tilde{\gamma}}$ are generated, yielding
\begin{equation}
\barr{rcl}
\mathcal{O}^e_s(\text{eff.})&=&\dis 
\frac{\widehat{\mathcal{C}_{\gamma}^e}}{\Lambda^2}
\, \varphi^*\varphi  \,\bar{e}e
\\[3ex]
\mathcal{O}^e_p(\text{eff.})&=&\dis
\frac{\widehat{\mathcal{C}_{\tilde{\gamma}}^e}}{\Lambda^2}
 \,\varphi^*\varphi \,
       \bar{e}\gamma_5 e \ ,
\earr
   \label{one_loop_elec}
\end{equation}
where the Wilson coefficients 
$\widehat{\mathcal{C}_{\gamma, \tilde\gamma}^e}$(for
$\mathcal{C}_{\gamma,\tilde{\gamma}}=1$) are both given by (see
Appendix B for details)
\begin{equation}
\widehat{\mathcal{C}_{\gamma,\tilde{\gamma}}^e}
= \frac{-3m_e\alpha_{\rm em}}{\pi} \, 
			    \ln\frac{\Lambda^2}{q^2} \ .
\label{one_loop_elec_Wilson}
\end{equation}
The loop has to be calculated using a gauge-invariant regularization
procedure such as Pauli-Villars or the dimensional (calculating in $4
- \epsilon$ dimensions) method and we choose to use the latter. The
ubiquitous factor of $(2/\epsilon - \gamma_E)$ has been traded, as
usual, for the logarithmic factor. Note that, compared to the
electron-DM operators in eq.(\ref{the_operators}), these have an extra
factor of $m_e / \Lambda$. The overall factor of $\Lambda^{-2}$ is, of
course, a legacy of the ``tree-level'' $\varphi \varphi^* \gamma\gamma$
parent. The factor of $m_e$, on the other hand, appears as a chirality
flip is essential for the fermions to couple to the (scalar) DM.  The
Wilson coefficients, understandably, are dependent on the momentum
transfer $q$. It has been argued~\cite{Essig:2011nj} that the
appropriate scale is that operative for atomic transitions, namely
$q=\alpha_{\rm em} m_e$.  Note that such a choice also serves to
enhance the sensitivity of these experiments. The calculation of the
elastic WIMP-electron cross section is now straightforward and yields
\begin{equation}
  \sigma_{\varphi e}(\mathcal{O}_{\gamma})
   = \frac{(\widehat{\mathcal{C}_{\gamma}^e}\, m_e)^2}
          {4\pi\Lambda^4 (m_{\varphi}+m_e)^2}
 \ ,
\end{equation}
and similarly for $\mathcal{O}_{\tilde\gamma}$. 

\begin{figure}[H]
\centering
	  \includegraphics[width=70mm]{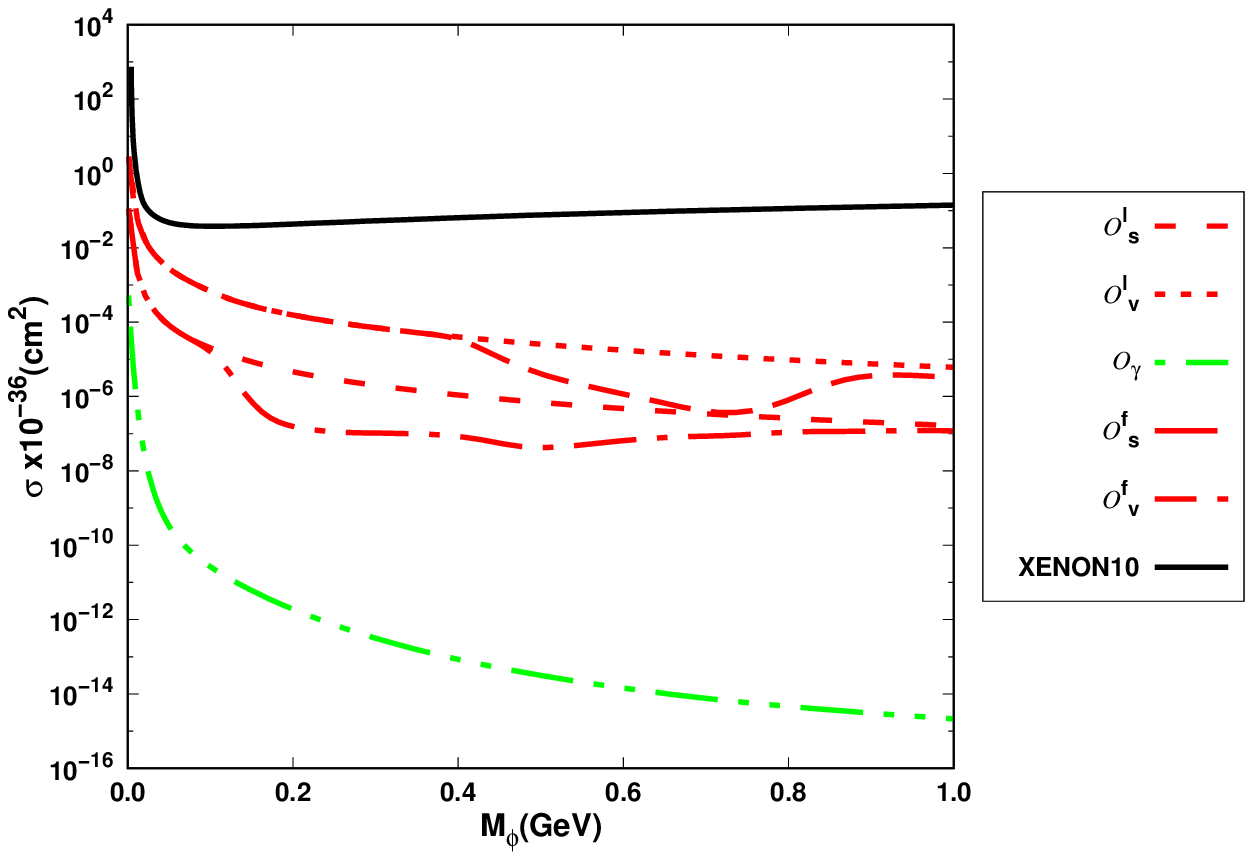} 
          \hspace*{10pt}
	  \includegraphics[width=70mm]{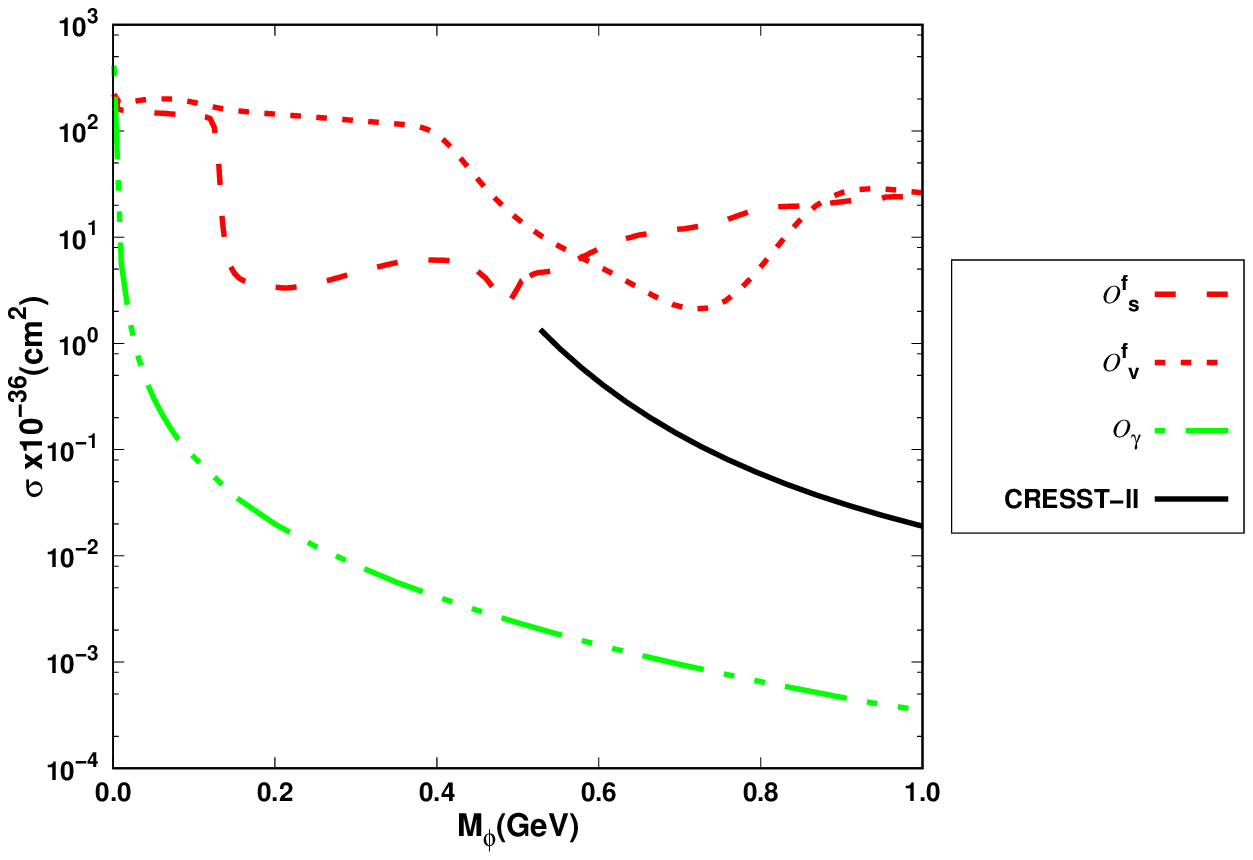}
  \caption{\small \em Scattering cross-sections of DM on {\em (a)} free electron
      and {\em (b)} nucleon $\sigma\left(\varphi+ e^-/N \to \varphi+ e^-/N
      \right)$ are depicted. The couplings are chosen are those that lead
      to the correct relic density. Exclusion plots from
      XENON10 and CRESST-II at 90\% C.L. for the case of {\em (a)}
      DM-electron scattering and {\em (b)} DM-nucleon scattering are also
      shown.}
\label{fig:dd} 
\end{figure}
Using values of $(m_\varphi, \Lambda)$ that {\em reproduce exactly} the Planck
measurements of the relic density (see Fig.5 of \cite{Choudhury:2019tss}),
we present, in Fig~\ref{fig:dd}$(a)$, the DM-electron elastic cross sections.  Also
presented, for comparison, are the XENON10 results. It might seem
paradoxical that the cross sections for ${\cal O}^f_{v}$ are larger
than those for ${\cal O}^f_{s}$ despite the $p$-wave suppression (see
eqn.~\ref{dd_eDM}). However, note that $\Lambda({\cal O}^f_{v}) \ll
\Lambda({\cal O}^f_{s})$, owing to the corresponding $p$-wave suppression in
DM-annihilation.  This more than makes up for the extra factor of
$m_\varphi^2/\Lambda^2$ in eqn.(\ref{dd_eDM}).

Note that we have used $\Lambda$-values corresponding to the case of
the democratic coupling. Had we considered a leptophilic DM instead,
the value of $\Lambda_{\rm max}$ for each of ${\cal O}^f_{s,v}$
would have been smaller by a factor ranging between 1.5--2 for the
$m_\varphi$ values of interest. This would translate to an increase in
the cross sections. However, as Fig.\ref{fig:dd} shows, the
cross-sections for ${\cal O}^f_{s}$ would still continue to be below
the XENON10 level.  On the other hand, those for ${\cal O}^f_{v}$
would start being comparable to the experimental upper bounds; in
particular, the range $\sim$5--150 MeV is already ruled out for such a theory.  It should also be
remembered that while the relic density constraint imposes an upper
bound on $\Lambda$, the direct detection experiments impose a lower
one. Thus, with only a little improvement, these experiments,
would start to rule out the parameter space allowed
by relic density.

A caveat needs to be entered here. In calculating the effective
  Wilson coefficients $\widehat{\mathcal{C}_{\gamma,
      \tilde\gamma}^e}$, we made two key assumptions in choosing the
  cutoff scale and in setting the momentum scale. Both the particular
  choices served to maximize $\widehat{\mathcal{C}_{\gamma,
      \tilde\gamma}^e}$ while remaining within the ambit of effective
  field theories. Uncertainties in these scales (related, as they are,
  to the ultraviolet completion) can significantly relax the bounds
  obtained from the non-observation of any signal at XENON10
.

\subsection{DM scattering off nucleons} For a DM with a mass greater than $0.5$ GeV, the parameter space can
be constrained using the negative results of the CRESST-II experiment.
To this end, we begin by evaluating the nuclear matrix elements
$\langle N|\mathcal{O}^f|N\rangle$ for all the operators at a scale
$\mu=$1 GeV. This leads to
\begin{equation}
\barr{rcl}
  \sigma_{\varphi N}(\mathcal{O}^f_{s})&=&\dis
\frac{F_{s,N}^2 m_N^2}{4\pi\Lambda^2 (m_{\varphi}+m_N)^2}
\\[3ex]
  \sigma_{\varphi N}(\mathcal{O}^f_{v})&=&\dis
\frac{F_{v,N}^2 m_N^2 m_{\varphi}^2}{\pi\Lambda^4 (m_{\varphi}+m_N)^2}
\earr
\end{equation}
where $m_N$ is the mass of the nucleon. The induced coupling constants
$F_{s,N}$ and $F_{v,N}$ parametrizing the effective DM-nucleon
interaction are given by 
\begin{eqnarray}
 F_{s,N}&=&\mathcal{C}^f_{s}\,\sum_{\text{all quarks}}F_{s,q}
              =\mathcal{C}^f_{s}\left(\sum_{\text{all quarks}}f_q^N\,\frac{m_N}{m_q}\right)\\
 F_{v,N}&=&3\,\mathcal{C}^f_{v}
\end{eqnarray}
where $m_q$ denotes the mass of quark, while $f_q^N$ are the proton
form factors. The latter can be calculated within various different
frameworks, with chiral perturbation theory giving some of the best
results.  For example, the proton form factors are $f_u^p=0.017$,
$f_d^p=0.036$, $f_s^p=0.043$,
$f_{c,b,t}^p= (2/27)\,(1-\sum_{q=u,d,s}f^p_q)=0.067$ as calculated
at $\mu=$1GeV~\cite{Crivellin:2013ipa}.

As for interactions mediated by a photon(s), the DM can scatter off a
single nucleon as well as an entire
nucleus~\cite{Weiner:2012cb,Frandsen:2012db}.  Let us begin by
focussing on the former, especially on the DM-nucleon interaction
generated on account of a DM-photon vertex.  At the one loop level,
the effective operators $\mathcal{O}^q_{s,p}(\text{eff.})$ would be
generated, just as in the case of the electrons (see
eq.\ref{one_loop_elec}).  The expressions for the corresponding Wilson
coefficients $\widehat{\mathcal{C}_{\gamma,\tilde{\gamma}}^q}$ would
be exactly identical to those in eq.(\ref{one_loop_elec_Wilson}),
apart from a multiplicative factor of $Q_q^2$ where $Q_q$ is the charge
of the quark under consideration. Quite apart from this, the DM can
also interact with the entire nucleus via the exchange of two virtual
photons\cite{Frandsen:2012db}. This coherent scattering implies that
the (two-photon) amplitude must scale as $Z^2$ where $Z$ is the atomic
number of the nucleus under consideration.  Scaling down this
amplitude by a factor of $A$ (the atomic weight) would then give us
the {\em average nucleon-DM} amplitude. Using the results of
Ref.\cite{Frandsen:2012db}, the induced operator for DM-nucleon
Rayleigh scattering can, then, be parametrized as
\begin{eqnarray}
\mathcal{O}_{Ray}&\simeq& 2 \sqrt{\frac{2}{\pi}} \, \frac{{\cal C}_{\gamma}~\alpha_{em} Z^2 Q_0}{A\Lambda^2} \;
F_\text{Ray}(\bar q) \, \varphi^*\varphi \, \bar{N}N \ ,
\end{eqnarray}
where $\bar q \equiv q / Q_0$ with $q$ being the momentum transfer and
the nuclear coherence scale $Q_0\sim 0.48(0.3+0.89A^{1/3})^{-1} \gev$.
The function $F_\text{Ray}(\bar q)$ is defined as~\cite{Frandsen:2012db}
\begin{equation} \label{eq:Fbarq}
\begin{split}
F_\text{Ray}(\bar q) & = -\sqrt{\frac{2}{\pi}} \int_0^1 \! d x \int_0^\infty d l \; \frac{l^2}{\left(l^2 + \bar q^2 (1-x)x \right)^2} \,  \exp \left[ - 2 \left ( l^2 - \bar q^2 \left ( (1-x) x - \tfrac{1}{2} \right ) \right ) \right] 
\\[2mm] & \phantom{xx} \times \left[ {\rm cosh} \left(2 \hspace{0.25mm} l \bar q \left (1-2x \right )\right) - \frac{ l^2 - \bar q^2 (1-x)x +\tfrac{1}{2}}{ l \bar q\left  (1-2x \right )} \; {\rm sinh} \left(2 \hspace{0.25mm} l \bar q \left (1-2x \right ) \right)\right] \, .
\end{split}
\end{equation}
The two amplitudes---those due to $\mathcal{O}^q_{s}(\text{eff.})$
and $\mathcal{O}_{Ray}$---add coherently, and, together, yield
\begin{eqnarray}
  \sigma_{\varphi N}(\mathcal{O}_{\gamma})&=&\frac{F_{{\gamma},N}^2 m_N^2}{4\pi \Lambda^4(m_{\varphi}+m_N)^2} 
\end{eqnarray} 
where
\begin{equation}
 F_{\gamma,N}=\left(\sum_{q=u,d,s}f_q^p\,\frac{m_N}{m_q}+\sum_{q=c,b,t}f_q^p\,\frac{m_N}{m_q}\right) 
 \widehat{\mathcal{C}_{\gamma,\tilde{\gamma}}^q} +  2 \sqrt{\frac{2}{\pi}} \, \frac{{\cal C}_{\gamma}~\alpha_{em} Z^2 Q_0}{A} \hspace{0.25mm} F_\text{Ray} \ .
\end{equation}

The CRESST-II experiment uses cryogenic detectors to search for
nuclear recoil events induced by the elastic scattering of dark matter
particles in calcium tungstate crystals. With the DM-nucleus
scattering cross section being proportional to $m_{A}^2$, the naive
expectation is that the dominant contribution to the scattering off a
$CaWO_4$ molecule would be that due to the tungsten nucleus. On the
other hand, the energy transferred in a scattering event is
approximately $q^2/(2m_A)$, where the transferred momentum $q
  \approx m_\varphi \, v_\varphi$. For large $m_A$, this would fall below the
detector threshold energy, which, in this case is $\approx 307$~eV.
Consequently, a large fraction of the events corresponding to
scattering off tungsten and a slightly smaller (yet large) fraction
of those off calcium nuclei would not register. Acting in concert
with this is the fact of there being four times as many oxygen nuclei
as the others. Thus, using $Z = 8$ in the formulae above is a very
good approximation and is in very good agreement with the simulations
for DM-nucleus scattering given in Fig. 7 of
ref~\cite{Angloher:2015ewa}.

The consequent size of the Rayleigh scattering contribution to the
amplitude is smaller as compared to that induced by
$\mathcal{O}^q_s(\text{eff.})$. Moreover, they interfere
destructively. Using values of $(m_\varphi, \Lambda)$ that satisfy the
relic density measurements (Fig.5 of \cite{Choudhury:2019tss}),
we present the DM-nucleon
elastic cross sections for all the operators in Fig~\ref{fig:dd}$(b)$.
As is immediately apparent, constraints from the CRESST-II results are
very weak for $m_\varphi \lapp 400 \mev$ and essentially
irrelevant. This is but a reflection of the fact that such light DM
particles cannot transfer sufficient energy to the nucleon for the
event to register. Moreover, even for $m_\varphi \gapp 400 \mev$, the
operators ${\cal O}_{\gamma, \tilde\gamma}$ continue to escape the
bounds from CRESST-II. So would be the case for fermionic operators
that do not involve the light quark fields (note that the DM need not
be hadrophobic, {\em per se}). And, once again, caveats such as that
in the preceding subsection do hold; in fact, even more so on account
of uncertainties in the calculation of hadronic matrix elements at
such low momentum transfers.

\section{Conclusion}
   \label{sec:conclusion}
In this, the second of a two-part investigation of the interactions of
a light (MeV scale) scalar DM particle with the SM sector within the
framework of an effective field theory, we effect a systematic study of
the sensitivity of existing experiments of a varied hue to such a DM
particle. In this, we were guided by our analysis---presented in the
first paper~\cite{Choudhury:2019tss}---of the cosmological
constraints. Encompassing not only relic density constraints but also
those from the requirements that the annihilation of the DM does not
significantly alter either the ratio of the neutrino and photon
temperatures or the shape of the CMB spectrum, these calculations took
cognizance of the fact that sub-GeV DM particles, on annihilation into
colored fields, can only manifest themselves in bound states rather
than, say, quasi-free quarks. Considerations such as these, on the
inclusion of higher-order effects in bound-state dynamics, result in
a non-trivial shape of the allowed parameter space of the EFT, and
this is what we have considered here.

With the LHC being unsuited to the investigation of such light states,
we take recourse to the clean environment of the high-luminosity KEK-B
accelerator that is already under operation. A DM particle with
unsuppressed couplings to the electron or to the photon can be looked
for at an $e^\pm$ collider. As the DM particles can only be produced
in pairs, and as there must be at least one visible particle in the
final state, the simplest process is $e^-e^+ \to \varphi \varphi^*
\gamma$ {\em i.e.}, a photon accompanied by missing energy-momentum.
In fact, at a low-energy facility such as the Belle-II, this is indeed
the most sensitive channel\footnote{The bounds derived in
  ref.\cite{Yu:2016pqp} using the $e^+ e^- \to J/\psi + \mbox{missing
    energy}$ channel are indeed much stronger. Note, however, that the
  authors of that article posit a much stronger
  interaction of the DM particle with the charm quark than with the
  electron. Consequently, their constraints are not directly
  applicable to the present situation.}.  Analysing the sensitivity of
different DM-SM interactions through the two-dimensional differential
kinematic distributions corresponding to this channel, we find that
most of the parameter space allowed by relic density can be probed at
Belle-II. Furthermore, based on a $\chi^2$ comparison of the
one-dimensional normalized differential kinematic distributions (given
in Fig.\ref{fig:distribution}) corresponding to an integrated
luminosity of 1--50~${\rm ab}^{-1}$, we find the
sensitivity to be robust and pronounced.

While the sensitivity limits obtained here are independent of the
details of the ultraviolet completion, they are certainly dependent on
the tensorial structure of the effective current-current
interaction. In particular, in the event of a positive signature, the
phase space distributions would distinguish between different
interaction Lagrangians. Moreover, not only can Belle-II discover
thermal DM candidates in the parameter range that is allowed as of
now, it can also access parameter space that is not amenable to a
thermal DM explanation. In this sense, a discovery by Belle can, in
principle, shed light on the mechanism of DM production and sustenance
in the early universe.

Not limiting ourselves to the mono-photon channel alone, we also
examine other final states namely $e^-e^+ \to \varphi \varphi^* \ell
  \bar{\ell}$, where $\ell$ is any of electrons and muons. Owing to
the smaller cross-section, the sensitivity is statistics-limited, and
is weaker than that available to the mono-photon channel,
especially for low luminosities. However, once the design integrated
luminosity is reached, these additional channels are not much worse
off. More interestingly, in the event of a discovery, these
complementary channels could be very useful in unravelling the
tensorial structure of the couplings and, hence, act as a pointer to
the UV completion.

Each of the final states discussed above could also have been
investigated at the LEP. Looking at the published data and
archival analyses, we find that while the LEP studies in the
mono-photon channel could be easily reinterpreted in terms of the EFT
parameters, the constraints so derivable were much weaker than those
obtainable at Belle-II. The main contributing factor, of course, is
the much higher luminosity at the KEK-B.

Finally, in a fashion similar to collider experiments, direct
detection experiments such as XENON100 or CRESST-II can also be used
to constrain the effective Lagrangian for the DM. While a naive
reading of the negative results at the latter experiment would seem to
suggest that a DM governed by this effective Lagrangian and
reproducing the correct relic density must satisfy 
$m_{\varphi} \lapp 0.5 \gev$,  this conclusion too can be evaded, for example, if the DM were
hadrophobic. On the other hand, the channel at the Belle-II that we
propose would be unaffected by such an assumption.  In other words,
such experiments offer a welcome complementarity of sensitivities.

{\bf Acknowledgements} DC acknowledges partial support from the
  European Union’s Horizon 2020 research and innovation program under
  Marie Sk{\l}odowska-Curie grant No 690575. DS thanks UGC-CSIR, India
  for financial assistance. 
\appendix
\section{Appendix A}

\begin{figure}[!htb]
\begin{center}
\includegraphics{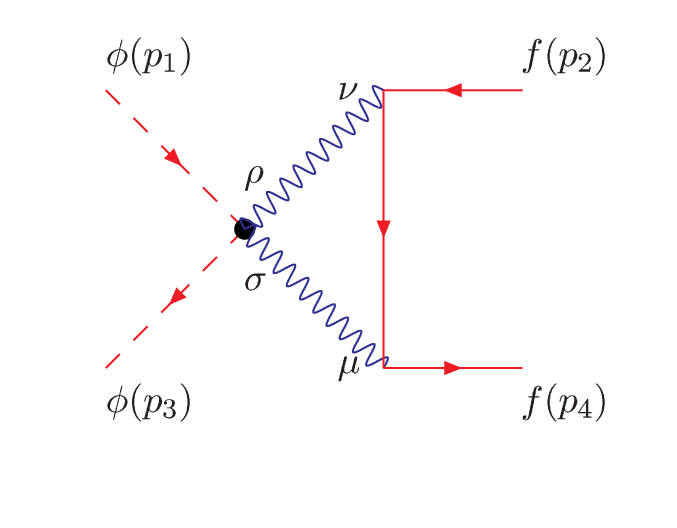} 
\end{center}
\vskip -30pt
\caption{\small \em{Typical Feynman diagram generating a $\varphi^* \varphi \bar f f$ coupling from ``tree-level''  $\varphi^* \varphi \gamma \gamma$ coupling.}}
\label{fig:feyn} 
\end{figure}

We consider here the effective $\varphi^* \varphi \bar f f$ coupling induced,
at the one-loop level, by a ``tree-level''  $\varphi^* \varphi \gamma \gamma$
coupling. The corresponding diagram is given in Fig.\ref{fig:feyn}.

\noindent
Denoting $q=p_1-p_3=p_4-p_2$,  the photon momenta in the loop above are 
$q/2 \pm k$ and the integral is
\begin{equation}
\barr{rcl}
 i\mathcal{M}&=& \dis \frac{4~i}{\Lambda^2} \,
     \int \frac{d^4k}{(2\pi)^4} \, \bar{u}_4 \, (-i \, e \, q_f \, \gamma_\mu)\,
         \frac{i}{\slashed p_{4}-\slashed{q}/2+\slashed{k}-m} \, 
 (-i \, e \, q_f \, \gamma_\nu)u_2 \, 
     \frac{(-ig^{\nu\rho})}{(q/2+k)^2}\frac{-ig^{\mu\sigma}}{(q/2-k)^2}  
\\[2ex]
 && \dis \hspace*{5em} 
   \left[\left(\frac{q}{2}+k\right)_\sigma\left(\frac{q}{2}-k\right)_\rho-\left(\frac{q^2}{4}-k^2\right)g_{\rho\sigma}\right] 
\\[3ex]
 &=& \dis \frac{-4\,e^2 \, q_f^2}{\Lambda^2} \,
    \int\frac{d^4k}{(2\pi)^4} \, 
        \frac{\bar{u}_4 \, \gamma_\sigma \, (\slashed{p}_{4}-\slashed{q}/2
             +\slashed{k}+m) \, \gamma_\rho \, u_2}
             {\left[(p_4-q/2+k)^2-m^2\right] (q/2+k)^2(q/2-k)^2}
 \\[2ex]
&& \dis  \hspace*{5em} 
\left[\left(\frac{q}{2}+k\right)_\sigma\left(\frac{q}{2}-k\right)_\rho
          -\left(\frac{q^2}{4}-k^2\right)g_{\rho\sigma}\right] \ ,
\earr
 \label{eqn:M}
\end{equation}
where we have used ${\cal C}_\gamma = 1$ and ${\cal C}_{\tilde\gamma} = 0$. 
Using Feynman parametrization, we may write
\begin{eqnarray*}
 \left\{
 \left[\left(p_4-\frac{q}{2}+k\right)^2-m^2 \right] \,
    \left(\frac{q}{2}+k\right)^2 \, \left(\frac{q}{2}-k\right)^2 
\right\}^{-1}
&=& \dis 2\int_0^1dx\int_0^{(1-x)}dy \, \frac{1}{\mathcal{D}^3}
\end{eqnarray*}
with
\[
\barr{rcl}
\mathcal{D}&=&\dis  x \, \left(\frac{q}{2}+k \right)^2
              +y\, \left(\frac{q}{2}-k\right)^2
              +(1-x-y)\, \left[ \left(p_4-\frac{q}{2}+k\right)^2-m^2\right]
\\[2ex]
& = & (k + a)^2 - \Delta
\\[3ex]
a_\mu & \equiv & \dis k_\mu+(1-x-y) \, p_{4\mu} + \frac{2x-1}{2} \, q_\mu
\\[2ex]
\Delta  & \equiv & \dis (1-x-y) \, q \cdot p_4 - a_\mu a^\mu \ .
\earr
\]
For the sake of simplicity,  we have set the external fermions
to be on-shell ($p_2^2 = p_4^2 = m^2$).
 Thus, the integral can be written as
\begin{equation}
\barr{rcl}
 \mathcal{M} & = & \dis \frac{8\,i\,e^2\,q_f^2}{\Lambda^2} \,
         \int_0^1dx\int_0^{(1-x)}dy\int\frac{d^4k}{(2\pi)^4} \,
        \frac{\mathcal{N}_1+\mathcal{N}_2}{\left[(k+a)^2-\Delta\right]^3}
\\[3ex]
 \mathcal{N}_1&\equiv& \dis 
   \left[\bar{u}_4 \,\gamma^\alpha \left\{\gamma^\beta \, 
                            \left(p_4-\frac{q}{2}+k\right)_\beta
                            + m \right\} \gamma^\rho\, u_2 \right]\;
        \left(\frac{q}{2}+k\right)_\alpha \, \left(\frac{q}{2}-k\right)_\rho
    \\[3ex]
\mathcal{N}_2&=&\dis 2 \, \bar{u}_4 \,
          \left(\slashed{p}_{4}- \,\frac{\slashed{q}}{2}+\slashed{k}+m
          \right)\,u_2 \; \left(\frac{q^2}{4}-k^2\right)
\\[2ex]
 &=& \dis 2\, \left(\frac{q^2}{4}-k^2\right) \; 
        \bar{u}_4 \,\left(\slashed{k}+2m \right) \, u_2
\earr
\end{equation}
Quite expectedly, the integral is quadratically
divergent. Consequently, we shall effect a dimensional regularization,
working in $d \, (=4-\epsilon)$ dimensions. We can, then, effect a 
Wick-rotation ($k_\mu \to k_{E\mu} \equiv l_\mu$) followed by 
a shifting of the integration variable. This leads to 
\begin{equation*}
 \mathcal{M}=\frac{8\,e^2\,q_f^2}{\Lambda^2} \, 
     \int_0^1dx\int_0^{(1-x)}dy\int\frac{d^dl}{(2\pi)^d}\frac{\mathcal{N}_1 +\mathcal{N}_2 }{(l^2+\Delta)^3}
\end{equation*}
Retaining only those terms in the numerator that lead to divergent terms, we have
\begin{equation*}
 \barr{rcl}
  \mathcal{N}_1
 &{\longrightarrow}&\dis \left(\frac{2+d}{d}(1-x-y)-\frac{2}{d}\right) \, m \, l^2 \, 
          \left[\bar{u}_4 \, u_2 \right] + {\rm finite}
\\[2ex]
 \mathcal{N}_2
 &{\longrightarrow}&\dis 2 \, \left(\frac{(2+d)}{d}(1-x-y)-2\right) \, m \, l^2 \,
        \left[\bar{u}_4 \, u_2 \right]  + {\rm finite}
\earr
 \end{equation*}
Finally, we have, 
\[
\mathcal{M} = \frac{-3 \, m \, q_f^2 \, \alpha_{\rm em}}{\pi \, \Lambda^2} \, 
       \left(\frac{2}{\epsilon}-\gamma_E\right) \, 
    \left[\bar{u}_4 \, u_2 \right] +{\rm finite}.
\]
In other words, a quantum of the operator ${\cal O}_s^f$ is generated. 
Here the factor of $m$ can be understood from the need to effect 
a chirality flip. It should be noted that even ${\cal O}_v^f$ would 
be generated, and without the factor of $m$. However, this would 
arise only from the finite pieces of the integral and, hence, 
would be suppressed. Effecting the usual replacement for the 
$(2/\epsilon - \gamma_E)$ factor, we finally have
\begin{equation}
{\cal C}_s^f({\rm eff.}) = \frac{-3 \, m \, q_f^2 \, \alpha_{\rm em}}{\pi \, \Lambda^2} \, 
               \ln \frac{\Lambda^2}{\mu^2} 
\end{equation}
where $\mu$ is the momentum scale of interest. 

Had we started from ${\cal O}_{\tilde \gamma}$ with ${\cal C}_{\tilde \gamma} = 1$ instead, 
we would have had
\begin{equation}
{\cal C}_p^f({\rm eff.}) = \frac{-3 \, m \, q_f^2 \, \alpha_{\rm em}}{\pi \, \Lambda^2} \, 
               \ln \frac{\Lambda^2}{\mu^2} 
\end{equation}
Once again, a suppressed but nonzero ${\cal C}_a^f({\rm eff.})$ would also be
generated.

In a similar vein, starting with the fermion operators, one 
could, analogously, generate ${\cal C}_{\gamma, \tilde\gamma}({\rm eff.})$ as well.

\bibliographystyle{JHEP}
\bibliography{reference}

\end{document}